\documentclass[desactivate]{aa}  
\usepackage{graphicx}
\usepackage{xcolor}
\usepackage{txfonts}
\usepackage{hyperref}

\begin{document}

   \title{The completeness of the open cluster census towards the Galactic anticentre}

   \author{Emily L. Hunt
          \inst{1,2}
          \and
          Tristan Cantat-Gaudin\inst{2}
          \and
          Friedrich Anders\inst{3,4,5}
          \and
          Lorenzo Spina\inst{6}
          \and
          Lorenzo Cavallo\inst{7}
          \and
          Alfred Castro-Ginard\inst{3,4,5,8}
          \and
          Vasily Belokurov\inst{9}
          \and
          Anthony G. A. Brown\inst{8}
          \and
          Andrew R. Casey\inst{10,11,12}
          \and
          Ronald Drimmel\inst{13}
          \and
          Morgan Fouesneau\inst{2}
          \and
          Sabine Reffert\inst{1}
          }

   \institute{
            Landessternwarte, Zentrum f\"ur Astronomie der Universit\"at Heidelberg, K\"onigstuhl 12, 69117 Heidelberg, Germany\\
            \email{emhunt@mpia.de}
        \and 
            Max-Planck-Institut f\"ur Astronomie, K\"onigstuhl 17, 69117 Heidelberg, Germany
        \and 
            Departament de Física Quàntica i Astrofísica (FQA), Universitat de Barcelona (UB), Martí i Franquès, 1, 08028 Barcelona, Spain
        \and 
            Institut de Ciències del Cosmos (ICCUB), Universitat de Barcelona (UB), Martí i Franquès, 1, 08028 Barcelona, Spain
        \and 
            Institut d'Estudis Espacials de Catalunya (IEEC), Edifici RDIT, Campus UPC, 08860 Castelldefels (Barcelona), Spain
        \and 
            INAF, Osservatorio Astrofisico di Arcetri, Largo E. Fermi 5, 50125 Firenze, Italy
        \and 
            Dipartimento di Fisica e Astronomia, Università di Padova, Vicolo dell’Osservatorio 3, 35122 Padova, Italy
        \and 
            Leiden Observatory, Leiden University, Einsteinweg 55, 2333 CC Leiden, The Netherlands
        \and 
            Institute of Astronomy, University of Cambridge, Madingley Road, Cambridge CB3 0HA, UK
        \and 
            School of Physics and Astronomy, Monash University, Clayton, VIC 3800, Australia
        \and 
            Centre of Excellence for Astrophysics in Three Dimensions (ASTRO-3D), Melbourne, Victoria, Australia
        \and 
            Center for Computational Astrophysics, Flatiron Institute, 162 Fifth Ave, New York, NY 10010, USA
        \and 
            INAF, Osservatorio Astrofisico di Torino, Strada Osservatorio 20, Pino Torinese 10025, Torino, Italy
        }

   \date{Received 15 October 2024; accepted 12 June 2025}
 
  \abstract
   {Open clusters have long been used as tracers of Galactic structure. However, without a selection function to describe the completeness of the cluster census, it is difficult to quantitatively interpret their distribution.}
   {We create a method to empirically determine the selection function of a Galactic cluster catalogue. We test it by investigating the completeness of the cluster census in the outer Milky Way, where old and young clusters exhibit different spatial distributions.}
   {We develop a method to generate realistic mock clusters as a function of their parameters, in addition to accounting for \emph{Gaia}'s selection function and astrometric errors. We then inject mock clusters into \textit{Gaia} DR3 data, and attempt to recover them in a blind search using HDBSCAN.}
   {We find that the main parameters influencing cluster detectability are mass, extinction, and distance. Age also plays an important role, making older clusters harder to detect due to their fainter luminosity function. High proper motions also improve detectability. After correcting for these selection effects, we find that old clusters are $2.97\pm0.11$ times more common at a Galactocentric radius of 13~kpc than in the solar neighbourhood -- despite positive detection biases in their favour, such as hotter orbits or a higher scale height.}
   {The larger fraction of older clusters in the outer Galaxy cannot be explained by an observational bias, and must be a physical property of the Milky Way: young outer-disc clusters are not forming in the outer Galaxy, or at least not with sufficient masses to be identified as clusters in \emph{Gaia} DR3. We predict that in this region, more old clusters than young ones remain to be discovered. The current presence of old, massive outer-disc clusters could be explained by radial heating and migration, or alternatively by a lower cluster destruction rate in the anticentre.}


   \maketitle

\section{Introduction}\label{sec:introduction}

\begin{figure*}[t]
    \centering
    \includegraphics[width=0.99\textwidth]{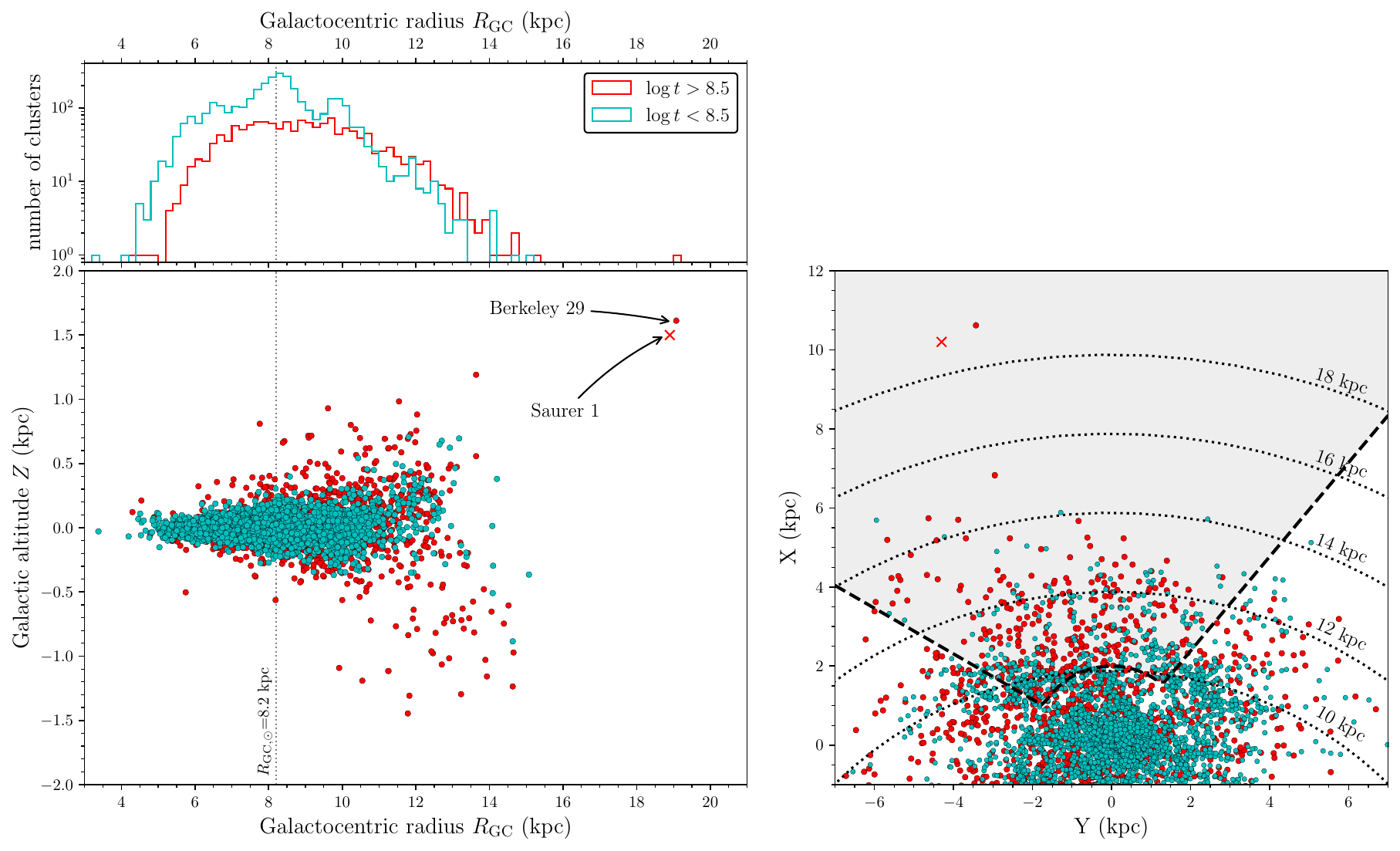} 
    \caption{\label{fig:XY_RgcZ} Spatial distribution of high-certainty clusters (\texttt{CST}$>$4) from \citet{2023A&A...673A.114H}. Ages are taken from \citet{2024AJ....167...12C} as they are more accurate for old clusters (see Sect.~\ref{sec:catalogues} for discussion).
    \emph{Top left}: Histogram of cluster galactocentric radii divided into young ($\log t < 8.5$, blue) and old ($\log t > 8.5$, red) clusters and with a 200~pc bin width. 
    \emph{Bottom left}: Distribution of the same young and old clusters but in terms of altitude $Z$ and Galactocentric radius $R_{\mathrm{GC}}$, assuming $R_{\mathrm{GC},\odot}$=8.2\,kpc. 
    \emph{Bottom right}: Projection of these clusters in heliocentric Galactic co-ordinates, with the Sun located at ($X$,$Y$)=(0,0). The dotted lines indicate Galactocentric radii from 10 to 18\,kpc. The shaded area is the region investigated in this study ($140^{\circ} \leq \ell \leq 240^{\circ}$ starting 2\,kpc from the Sun) . In both lower panels, the cross indicates the cluster Saurer~1, visible in \textit{Gaia} data but not recovered in the blind search of \citet{2023A&A...673A.114H}.}
\end{figure*}

Star clusters are groups of a dozen to hundreds of thousands of coeval stars, born from the collapse of the same parent molecular cloud \citep{2010ARAA..48..431P}. Stars in a cluster can remain gravitationally bound to their siblings for billions of years. Systematic searches for star clusters were a popular activity in early modern astronomy \citep[e.g.][]{1781cote.rept..227M,1786RSPT...76..457H,1789RSPT...79..212H}. Since their main parameters (in particular age and distance) can be estimated more reliably than for individual stars, they have long been used as tracers of the structure and overall properties of the Milky Way \citep[e.g.][]{1930LicOB..14..154T,1982ApJS...49..425J}. Open clusters (OCs) are the most common type of star cluster in the Milky Way, with typical ages no greater than a few hundred million years \citep{2022Univ....8..111C} and masses ranging from less than a hundred to no larger than a few thousand solar masses \citep{AlmeidaMonteiro_2023,2024A&A...686A..42H,AlmeidaMoitinho_2025}.

Catalogues of stellar proper motions have enabled the identification of OCs as co-moving groups \citep[][]{2003A&A...410..565A,2005A&A...440..403K,2007MNRAS.374..399F}, but we remark that until recently the vast majority of the objects listed in the most widely used cluster databases \citep{2002A&A...389..871D,2012ASSP...29...53N,2013A&A...558A..53K} were discovered as simple spatial overdensities in on-sky positions. The ESA \textit{Gaia} mission \citep{2016A&A...595A...1G} has transformed our ability to study stellar populations in the Milky Way \citep{2021ARA&A..59...59B}, and its second \citep{GaiaDR2} and third data releases \citep{GaiaDR3} have enabled the discovery of thousands of new OCs\footnote{We refer the reader to \citet{2022Univ....8..111C} and \citet{2023MNRAS.526.4107P} for in-depth reviews of the \textit{Gaia} contribution to the cluster census.} in the 5D space of positions, proper motions, and parallaxes, in addition to validation with \emph{Gaia}'s three-colour photometry \citep[e.g.][]{2018A&A...618A..59C, 2019A&A...624A.126C,2020A&A...635A..45C,2023A&A...673A.114H}.

While the most recent cluster catalogues feature unprecedented numbers of objects characterised to a high level of precision, we still have a poor understanding of the completeness of the cluster census. It feels intuitive that massive and nearby clusters are easier to detect than small and distant objects. All-sky blind searches performed on the \textit{Gaia} catalogue are also likely to be biased in favour of clusters with proper motions different to their background field, and to be more efficient at recovering clusters in sparser regions of the sky than in crowded fields.
The issue of completeness dropping with distance is often sidestepped in astronomy by restricting studies to a volume within which the authors consider the sample is near 100\% complete \citep[e.g.][]{2008ApJ...686..279K,2013A&A...558A..53K,2024A&A...686A..42H}. This approach is in general highly sub-optimal \citep{2021AJ....162..142R}. 

Rather than artificially restricting catalogues to create volume-complete samples, we should strive to construct their selection function: a quantitative description of completeness as a function of location and other relevant parameters. A selection function allows one to account for selection effects in the incomplete regime, thus enabling studies of the Milky Way on a much larger scale. Even assuming that catalogues are complete within a certain volume, the selection function is a critical piece of information to extend studies beyond the solar neighbourhood, or to compare the Milky Way to other galaxies in an unbiased way. To our knowledge, the only study attempting to establish a selection function for the cluster census was performed by \citet{2021A&A...645L...2A} for the \textit{Gaia}~DR2 cluster catalogue of \citet{2020A&A...640A...1C}, which itself exploited a completeness experiment conducted by \citet{2020A&A...635A..45C} using \citet{2018A&A...618A..93C} as a reference catalogue.

The present paper focuses on the spatial and age distribution of clusters in the outer Milky Way, in the direction of the Galactic anticentre ($140^{\circ} \leq l \leq 220^{\circ})$, as well as an additional region where the Galactic warp appears most strongly in the cluster population ($220^{\circ} < l \leq 240^{\circ})$. While young clusters (defined in this study as younger than $\log t$=8.5) are six times more numerous than old clusters within 11\,kpc of the Galactic centre, old clusters are twice as numerous beyond 11\,kpc. Figure~\ref{fig:XY_RgcZ} shows that the number of known young clusters decreases more quickly with Galactocentric distance than for older objects. The distant outer-disc clusters are also mostly found in the third Galactic quadrant. Without a quantitative selection function, we cannot know whether this distribution is a physical property of the Milky Way or the result of a biased cluster census.

In this study, we inject mock clusters into \textit{Gaia} data, and we attempt to recover them in a blind search akin to typical cluster searches. We aim to develop a method applicable to any cluster blind search in the \emph{Gaia} era, and demonstrate it in use on the Galactic anticentre. This paper is structured as follows. Section~\ref{sec:catalogues} outlines the choice of catalogue and cluster retrieval method that we aim to establish a selection function for. Next, Sect.~\ref{sec:methods} presents our approach to generating mock clusters and detecting them in the \emph{Gaia} data. Section~\ref{sec:results} presents the results of our search and the selection function we build from this experiment. Section~\ref{sec:discussion} discusses the consequences of the recovered completeness estimates on our understanding of Galactic structure as traced by OCs. Finally, Sect.~\ref{sec:conclusion} summaries our findings and provides a brief outlook to future work.

\section{Cluster sample}\label{sec:catalogues}

\begin{table}[t]
    \caption{\label{tab:catalogues}Number of clusters in our studied region in HR23 and HR24 given multiple possible cuts.}
    \centering
    \begin{tabular}{rcc}
        \hline\hline
        Sample & HR23 & HR24 \\
        \hline
Total & 856 & 856 \\
Type = \texttt{o} & 837 & 818\tablefootmark{a} \\
$Q_\text{CMD} > 0.5$ & 709 & 709 \\
Type = \texttt{o}, $Q_\text{CMD} > 0.5$ & 696 & 682 \\
CST > 5$\sigma$ & 613 & 613 \\
Type = \texttt{o}, CST > 5$\sigma$ & 600 & 603\tablefootmark{b} \\
Type = \texttt{o}, $Q_\text{CMD} > 0.5$, CST > 5$\sigma$ & 518 & 523 \\
        \hline
    \end{tabular}
    \tablefoot{\tablefoottext{a}{Low-quality sample adopted in this work.} \tablefoottext{b}{High-quality sample adopted in this work.}}
\end{table}

The current cluster census includes multiple different catalogues and individual cluster searches generated with \emph{Gaia} data releases, often using a wide range of different methodologies (see Introduction). For the purposes of this work, we aim to establish the selection function of OCs in the \cite{2024A&A...686A..42H} (hereafter HR24) catalogue, as it is a recently developed catalogue based on the latest \emph{Gaia} data release (DR3) that also includes parameters determined for all objects. Nevertheless, establishing the selection function of the HR24 catalogue is not without some complications that we discuss in this section. As HR24's cluster list is an updated version of the \cite{2023A&A...673A.114H} (hereafter HR23) catalogue, construction of the HR24 catalogue hence involved a number of steps that must be replicated to establish its full selection function, which we discuss below.

HR24 is based on a cluster search performed in HR23. HR23 applied the Hierarchical Density-Based Spatial Clustering of Applications with Noise \citep[HDBSCAN,][]{hdbscan_paper, McInnes2017} algorithm to recover star clusters from \emph{Gaia} DR3, recovering a total of 7167 clusters. To augment the HDBSCAN algorithm, as it often produces a high number of false positives when applied to \emph{Gaia} data, they used the cluster significance test (CST) developed in \cite{2021A&A...646A.104H}, which measures the statistical significance of a candidate cluster compared to its immediate surroundings. To remove unreliable clusters and likely false positives, HR23 includes only 7167 cluster candidates with $\text{CST}> 3\sigma$, which is the minimum bar that a cluster must pass to be included in HR23 or HR24.

Where the two catalogues differ is in their definition of an OC. HR23 found that HDBSCAN was sensitive to other types of star cluster than simply OCs, with their catalogue including over 100 globular clusters and a large number of clusters that appeared to be unbound moving groups, particularly in the region close to the Sun. Classification of globular clusters was performed with comparison to the literature. To classify OCs and moving groups, HR23 used the statistical cuts on cluster radius and proper motion dispersion defined in \cite{cantat-gaudin_clusters_mirages_2020}; however, they found that these cuts appeared to be insufficient for many more compact moving groups. Hence, HR24 introduced a method based on cluster mass and radius, classifying 1309 clusters from HR23 as moving groups. In practice, in our studied region in the anticentre and at distances greater than 2~kpc, the differences between the two catalogues are minor: only an additional 18 out of 856 clusters in our region are classified as moving groups in HR24, in addition to one extra globular cluster (see Table~\ref{tab:catalogues}). Almost all of these moving groups are removed with a more stringent cut on CST of $5\sigma$. For the purposes of this work, we use only the sample of clusters in HR24 classified as OCs (Type = \texttt{o}), although future work on selection functions near to the Sun (where the moving group fraction is higher) may also need to additionally calculate the probability that a true OC was correctly classified as one by the HR24 pipeline.

It is also worth noting that HR23 and HR24 include a `high quality' cluster sample. HR23 used a neural network to classify cluster colour-magnitude diagrams (CMDs) and assign them a quality $Q_\text{CMD}$ between zero and one, and they recommended some analyses instead use a more strongly cut high quality cluster sample: defined as clusters with $\text{CST}>5\sigma$ and $Q_\text{CMD}>0.5$. In this work, we do not establish the selection function of their high quality cluster sample, as this would be difficult to do fairly: given that their CMD classifier was trained on simulated clusters and then applied to real clusters, and was around three times more accurate at classifying simulated clusters than real clusters, it is likely that presenting their CMD classifier with our simulated clusters would strongly overestimate the probability that a real cluster has a good CMD class. Nevertheless, the selection function of the HR23 cluster classifier could still be worthwhile to explore in a future work.

Finally, it is also worth commenting on the sample of cluster ages that we adopt. \cite{2024AJ....167...12C} used a machine learning method to fit isochrones to clusters in HR23. They found that cluster ages can be underestimated for the oldest clusters in HR23 (and by extension HR24, which contains the same age parameters). For this work, as we are interested particularly in old clusters, we use only the \cite{2024AJ....167...12C} cluster ages and extinctions for HR24 clusters in our analysis and figures.

\section{Methods for creating an empirical OC selection function}\label{sec:methods}

In this section, we describe all steps in our methodology to calculate an empirical cluster selection function. Firstly, we describe how we simulated realistic clusters. Next, we describe the process of injecting and retrieving them in a manner similar to that in HR23. Finally, we discuss how we post-processed these results to create a smooth cluster selection function.

\subsection{Creation of simulated clusters}\label{sec:simulating_clusters}

To derive a robust selection function of the OC census using retrieval of simulated clusters, it is essential that our simulated clusters are as realistic as possible. Hence, the first step in our method was the development of a robust simulated cluster generation pipeline.

There are a number of parameters that must be chosen to create a \emph{Gaia} observation of a simulated cluster, which we discuss in the following sections. Some of these parameters are intrinsic (selected during this study), and some of them are extrinsic (such as properties of \emph{Gaia} observations). The intrinsic parameters of a simulated cluster are: its position, which we quantify in Galactocentric co-ordinates $l$ and $b$ as well as its distance from the Sun $d$; its total mass, $M$, with a corresponding number of stars $N$ sampled from an assumed mass function of the cluster; the structure of the cluster, defined by a \cite{1962AJ.....67..471K} profile; the current velocity of the cluster, defined as its proper motions $\mu_{\alpha^*}$ and $\mu_\delta$, as well as its radial velocity; the velocity dispersion of the cluster's member stars, $\sigma_\text{1D}$; the cluster's age, $\log t$; and its metallicity, $[\text{Fe}/\text{H}]$. The extrinsic parameters of a cluster are: the assumed mean extinction $A_V$ of the cluster; the distribution of \emph{Gaia} photometric and astrometric uncertainties at the cluster's position; the selection function of the \emph{Gaia} satellite and the subsample of \emph{Gaia} data analysed in HR23 at the cluster's position, causing many member stars to be missing from the analysed dataset; and the resolving power of \emph{Gaia}, which causes some multiple star systems to be resolved as a single source. Having selected intrinsic parameters for a given cluster, creation of simulated clusters involves two steps: firstly, generation of cluster member star photometry; and secondly, generation of cluster member star astrometry.

\subsubsection{Simulation of member star photometry}
\label{sec:simulating_photometry}

To generate mock photometry for a given cluster age, metallicity, and total mass, we began by sampling the masses of its member stars from a \citet{Kroupa2001_IMF} initial mass function using the Python package \texttt{IMF}\footnote{\url{https://github.com/keflavich/imf}}. We then assigned the stars an absolute \emph{Gaia} DR3 $G$, $G_{BP}$, and $G_{RP}$-band magnitude using version 1.2S PARSEC isochrones \citep{Bressan2012MNRAS.427..127B, Chen2014MNRAS.444.2525C, Tang2014MNRAS.445.4287T, Chen2015MNRAS.452.1068C}.

We randomly picked a location for each cluster, with a Galactic longitude $l$ between 140\,$^{\circ}$ and 240\,$^{\circ}$, Galactic latitude $b$ between -10\,$^{\circ}$ and +10\,$^{\circ}$, and a heliocentric distance $d$ between 2 and 15\,kpc, sampled uniformly within these ranges. We then assigned interstellar extinction to the cluster based on the value reported in the 3D dust map of \cite{2019ApJ...887...93G} at its location.

Having applied foreground extinction to our cluster member stars, we simulated additional extrinsic effects on them. Firstly, we simulated the chance that a star is in a binary system using the unresolved binary star correction code of HR24. In brief, this method takes all simulated stars for a given cluster, and iteratively pairs them into binaries based on primary mass-dependent binarity probabilities from \cite{2017ApJS..230...15M}. Then, the average separation between the host and orbiting star from \cite{2017ApJS..230...15M} was used to estimate whether \emph{Gaia} would resolve the system as a resolved or unresolved, and hence whether or not to `merge' the two sources into one. Some stars being in unresolved binary systems reduces the total number of sources observed by \emph{Gaia} for a given cluster, and hence the detectability of the overall cluster is reduced (especially for low-mass clusters). 

The next step was to simulate the probability that a given simulated star would be in the subsample of \emph{Gaia} data analysed by HR23. To estimate the probability that a simulated star would even appear in the \emph{Gaia} dataset initially, we compared its $G$-band magnitude to the \emph{Gaia} selection function of \cite{2023A&A...669A..55C} at the star's position and magnitude, and rejected the star if a random uniform variable is greater than the corresponding value of the \emph{Gaia} selection function. In practice, this removes many faint sources in our simulated clusters that are below \emph{Gaia}'s magnitude limit, which is dependent on position and can be as faint as $G \approx 21$.

However, HR23 analysed a subsample of \emph{Gaia} data after applying quality cuts; incorporating a similar quality cut on our simulated clusters is important to ensure that they have the same number of stars and luminosity function as a real cluster in HR23, particularly at the faint end of the cluster ($G \gtrsim 18$ mag) where HR23's quality cuts have the largest impact. HR23 performed clustering only on \emph{Gaia} sources with proper motions and parallaxes, as well as only those with $G$, $G_{BP}$, and $G_{RP}$ photometry; in addition, they cut sources with low-quality astrometry with a \cite{2022MNRAS.510.2597R} \texttt{v1} quality flag below 0.5. Simulating the reason why a given star may fail one or more of these cuts is complicated: reasons such as a source's multiplicity, crowding, brightness, or colour could cause \emph{Gaia}'s astrometric or photometric fitting to fail or produce a low-quality result \citep{2021A&A...649A...2L,2021A&A...649A...3R}. As an approximation, we calculated the subsample selection function of the HR23 cuts on \emph{Gaia} with the method outlined in \cite{2023A&A...677A..37C} and on a grid of HEALPix\footnote{\url{http://healpix.sourceforge.net}} \citep{2005ApJ...622..759G} level 6 pixels across our studied region in the Galactic anticentre, creating a selection function that depends on position and $G$-band magnitude. Then, for a given cluster, we applied the selection function in the HEALPix pixel they are in to the stars in the cluster based on their $G$-band magnitude. This approximation does not simulate when a certain combination of binary star orbital parameters causes a source to have low-quality astrometry with \emph{Gaia}; however, it does capture the impact of crowding on the \emph{Gaia} pipeline, and how faint sources often do not have proper motions and parallaxes -- hence causing them to fail the HR23 quality cuts, reducing the number of stars at the faint end of clusters in HR23. For a typical cluster in the anticentre, our approximate subsample selection function reduces the magnitude limit of our simulated clusters from $G\sim20.7$ to $G\sim19$.

For this study, we do not simulate further impacts on the photometry of our simulated clusters, such as \emph{Gaia} photometric uncertainties -- for similar reasons as to why we do not simulate differential reddening. We recall that our clustering algorithm only uses 5D astrometric measurements. Thus, we do not require fully realistic cluster CMDs, and only need stellar $G$-band magnitudes to assign astrometric errors to stars in the following section. The uncertainties on $G$-band magnitudes are small enough to be insignificant to this step.

\subsubsection{Simulation of member star astrometry}
\label{sec:simulating_astrometry}

To generate member star astrometry, we began by generating the 3D positions of stars in each cluster. We assumed that all clusters follow a King profile, parameterised by a core radius $r_c$ and a tidal radius $r_t$. To select $r_t$ for each cluster, we assumed that all clusters fill their Roche lobe, with their tidal radius equalling their Jacobi radius; we hence calculated the expected tidal radius of each cluster based on their mass and the potential of the Milky Way. For the potential, we used the \texttt{MWPotential2014} model in \texttt{galpy}\footnote{\url{http://github.com/jobovy/galpy}} \citep{Bovy15}, which models the potential of the Milky Way with a simple, smooth model, composed of a a bulge, disk, and dark matter halo. Then, the potential is used to calculate the cluster's Jacobi radius (Eqn. 2 in HR24). Given the weak cube root dependence of the cluster's Jacobi radius on the components of the potential model, the choice of potential has a relatively small impact on our study. Next, for each cluster, we chose a core radius of the cluster $r_c$ (defining how concentrated a given cluster is), sampling uniformly from a distribution chosen later in this work. Having selected values for $r_c$ and $r_t$, the 3D position of each star is then randomly sampled using the King profile sampler in \texttt{ocelot}\footnote{\url{https://github.com/emilyhunt/ocelot}} \citep{2021A&A...646A.104H}, which uses the equation for spatial density of a King profile \citep[Eqn. 27 in][]{1962AJ.....67..471K} and rejection sampling to sample random 3D co-ordinates given a King profile. These true positions were then converted to observed co-ordinates with parallaxes calculated as inverse distances by \texttt{Astropy} \citep{astropy:2013, astropy:2018, astropy:2022}.

Next, we generated the velocities of simulated cluster member stars, which begins by choosing an overall mean velocity for each cluster. In order to sample the full velocity space efficiently, we wish to sample all possible proper motions that clusters could reasonably have, which mostly includes proper motions corresponding to orbits reminiscent of stars in the thin disk where OCs mostly reside \citep{2022Univ....8..111C}, but could also hypothetically include higher-velocity evolved thick disk-like orbits. As a simple method to sample realistic proper motions, we used the proper motion distribution of field stars in each clustering region at a similar parallax to that of the cluster to derive cluster mean proper motions, assuming that field star proper motions follow a Gaussian distribution and selecting only those with a parallax within 0.1~mas of a simulated cluster's parallax. This method samples a wider range in proper motions than the proper motion distribution of known clusters (see Fig.~\ref{fig:proper_motions}), ensuring that we can later experiment with how more or less extreme cluster proper motions can improve cluster detectability. For this study, we assumed that clusters have no radial velocities, as the impact of radial velocities on proper motions \citep[perspective acceleration; ][]{Lindegren2021} is insignificant at the distances we consider in this work, as perspective acceleration only measurably impacts clusters that have a large on-sky distribution -- generally only those within a few hundred parsecs.

Having chosen a cluster mean proper motion, we generated the cluster's internal star-by-star dynamics by sampling the true 3D velocity of stars from a multivariate Gaussian distribution, with a dispersion given by the velocity dispersion equation in \cite{2010ARAA..48..431P} (hereafter PZ10) and assuming that clusters are in virial equilibrium (i.e. the constant $\eta$ in their equation equals ten). In practice, at the distances and cluster masses considered in this work, even for a highly supervirial cluster, the internal velocity dispersion within star clusters is smaller than \emph{Gaia} proper motion uncertainties -- for instance, a relatively high mass virialised $10^3$~M\textsubscript{\sun} Pleiades-like cluster at a distance of 5~kpc has an internal dispersion of just $\sim0.02$~mas\,yr\textsuperscript{-1}. Nevertheless, our pipeline includes internal cluster velocity dispersion in preparation for future work focusing on closer objects.

Finally, we sampled observed cluster member proper motions and parallaxes that include \emph{Gaia} measurement uncertainties.\emph{Gaia} astrometric uncertainties are principally a function of sky position and G-band magnitude (\citealt{2021A&A...649A...2L}, \citealt{Fabricius2021}). We used the data-driven approach of HR24 to calculate expected astrometric errors: in the field of analysed \emph{Gaia} data around each cluster in this study, the nearest real star in $G$-band magnitude to a simulated star was chosen, and its measurement uncertainties were assigned to the simulated star. In the future, an approach that also uses $G_{BP}$ and $G_{RP}$ photometry may be more accurate. Sampling from a Gaussian distribution, these astrometric uncertainties were then added on to the true astrometry of each simulated star to create its mock \emph{Gaia} observation.

\subsection{Cluster blind search}\label{sec:blind_search}

\subsubsection{Cluster recovery method}

\begin{figure*}[t]
    \centering
    \includegraphics[width=0.99\textwidth]{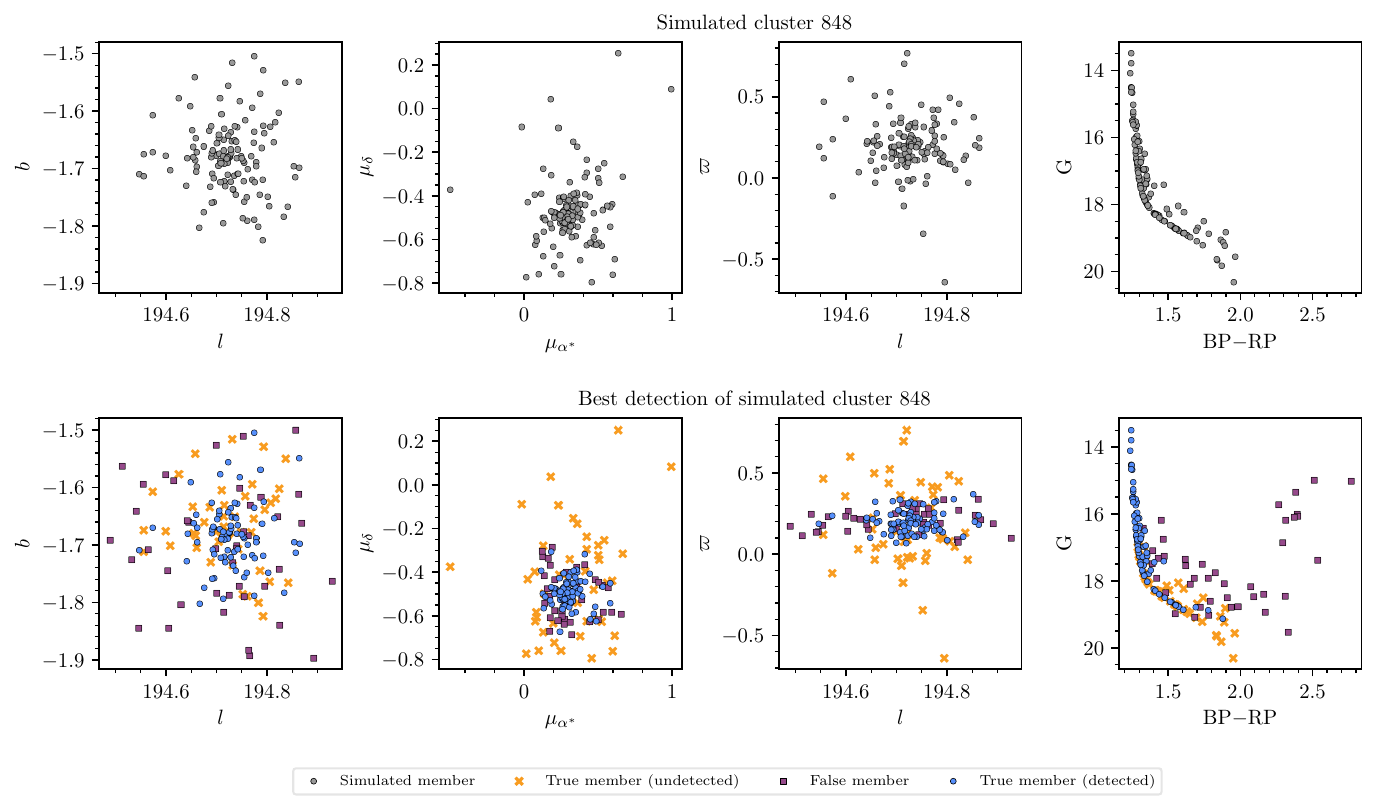} 
    \caption{\label{fig:simulated_cluster_detection} Example result of the injection-recovery procedure for a simulated OC, in the typical projections used in cluster discovery studies. \emph{Top row:} Original simulated cluster (including realistic {\it Gaia} DR3-like astrometric errors), shown in position (\emph{far left}), proper motion (\emph{centre left}), position versus parallax (\emph{centre right}), and the colour-magnitude diagram of the cluster (\emph{far right}). This simulated cluster was then inserted into \emph{Gaia} data to test the cluster recovery ability of HDBSCAN. \emph{Bottom row:} Same as top row, but for the best recovery from \emph{Gaia} data of this cluster by HDBSCAN. True members of the cluster that were correctly assigned as members by HDBSCAN are shown as blue circles. Non-member field stars from the \emph{Gaia} dataset that were incorrectly added to the cluster are shown as purple squares. Finally, member stars that HDBSCAN missed are shown as orange crosses.}
    
\end{figure*}

\begin{table}[t]
    \caption{\label{tab:parameter_choices}Sampled ranges in parameters for simulated clusters.}
    \centering
    \begin{tabular}{lcccc}
        \hline\hline
        Parameter & Source & Min & Max & Unit \\
        \hline
        $l$ & sampled & 140 & 240 & deg\\
        $b$ & sampled & -10 & +10 & deg\\
        $d$ & sampled & 2 & 15 & kpc\\
        $M$ & sampled & 50 & 5000 & M\textsubscript{\sun}\\
        $\log t$ & sampled & 6.4 & 10.0 & dex \\
        $[\text{M}/\text{H}]$ & fixed & -0.2 & -0.2 & dex \\
        $r_c$ & sampled & 1 & 5 & pc\\
        $r_t(R_{GC})$ & HR24 Eqn. 2 & 5.9 & 51 & pc \\
        $\mu_{\alpha*}(l,b,d)$ & field stars & -9.1 & 7.2 & mas\,yr\textsuperscript{-1} \\
        $\mu_{\delta}(l,b,d)$ & field stars & -9.8 & 11.2 & mas\,yr\textsuperscript{-1} \\
        $A_V(l,b,d)$ & \cite{2019ApJ...887...93G} & 0 & 13.4 & mag \\
        $\eta$ & fixed & 10 & 10 & - \\
        $\sigma_{1D}(M,r_{50},\eta)$ & PZ10 Eqn. 4 & 79 & 876 & m\,s\textsuperscript{-1} \\
        
        \hline
    \end{tabular}
    \tablefoot{Shown for intrinsic parameters, which are those sampled from a uniform distribution with a range chosen by the authors of this work; and for extrinsic parameters, which are cluster parameters that are derived as a function of chosen intrinsic parameters based on a model (see Sect.~\ref{sec:blind_search:parameters} for further description.)}
\end{table}

Using our simulated cluster generation pipeline, the next step in our method was to perform injection and retrieval of simulated clusters. We used the exact methodology from HR23 to retrieve the injected clusters, which we briefly outline here.

We began by dividing \emph{Gaia} data into HEALPix level five pixels in a Galactic co-ordinate frame, performing clustering on each level five pixel ($\sim${}3.4~deg\textsuperscript{2} in area) plus an overlap region of all eight neighbouring pixels surrounding it: hence, each clustering `field' contains nine HEALPix level five regions. As we aim to fully replicate the steps involved in the clustering in HR23, we also applied the HR23 quality cuts on \emph{Gaia} data, retaining only \emph{Gaia} sources with proper motions, parallaxes, and photometry in the $G$, $BP$, and $RP$ bands, in addition to cutting all stars with a \cite{2022MNRAS.510.2597R} \texttt{v1} quality flag below 0.5. We found that inserting four simulated clusters per field per run (corresponding to a density of around one cluster per square degree) maximised the number of clusters we could simulate injections and retrievals of, but without causing the cluster density to be so high that simulated clusters overlapping was common.

Having inserted simulated clusters into a region, we ran HDBSCAN \citep{hdbscan_paper,McInnes2017} at four different \texttt{min\_cluster\_size} values (10, 20, 40, and 80), and with \texttt{min\_samples} set to 10. Finally, for every detected cluster, we also computed its cluster significance test (CST) score \citep{2021A&A...646A.104H}, which measures the statistical significance of a cluster against the surrounding field as a way to remove false positive clusters -- effectively producing a signal to noise ratio of each cluster relative to the surrounding field. The HR23 study only retained clusters with $\text{CST} \geq 3\sigma$; hence, we consider any cluster with a CST below this to be undetected in HR23. However, HR23 also recommended using a higher cut when higher quality cluster samples are desired (such as $\text{CST} \geq 5\sigma$), and so we also investigate the selection function of some higher CST cuts in this work when analysis of a higher quality sample is required.

The use of increasing \texttt{min\_cluster\_size} values was adopted by HR23 to deal with the fact that HDBSCAN sometimes splits large clusters into multiple subclusters. For the present study, whenever a given \texttt{min\_cluster\_size} leads to none of the clusters being split into multiple components, we terminated the clustering early. This can result in as much as a 75\% speed-up in our methodology in cases in which all clusters would be detected (or non-detected) identically at all \texttt{min\_cluster\_size} values.

Figure~\ref{fig:simulated_cluster_detection} shows an example simulation and detection of a cluster, which includes realistic astrometric uncertainties and simulated \emph{Gaia} selection effects. The cluster is young ($\log t = 7.4$), has a total mass of 1412~M\textsubscript{$\sun$}, is at a distance of 5.2~kpc from the Sun, and has an extinction of $A_V=3.1$~mag. It should be noted that the CMD of the cluster is more ideal than for a real object: we did not simulate differential reddening and photometric uncertainties, as they do not affect our detection pipeline as it only uses astrometric data (see Sect.~\ref{sec:simulating_photometry}.) The simulated cluster contains 132 stars that would be visible in our subsample of the \emph{Gaia} dataset if this cluster was real. Of these stars, only 83 are assigned as members of the cluster, with many (predominantly faint) stars with higher astrometric uncertainties not being identified as cluster members. In addition, 46 field stars that are nearby, unassociated parts of the \emph{Gaia} dataset were erroneously assigned as cluster members. This example corresponds to a clear detection, with a CST of 16.2.

\subsubsection{Explored cluster parameter space}\label{sec:blind_search:parameters}

Having defined a cluster simulation and recovery pipeline, our final step was to decide on clusters to simulate and run our full injection and retrieval experiment. We investigated ranges in cluster parameters that fully sample the range of observed cluster parameters in the Galactic anticentre, which are summarised in Table~\ref{tab:parameter_choices}. For the purposes of this work, we fixed some parameters that have a minimal impact on our cluster detectability to reduce the size of our parameter space. Namely, we set cluster metallicity $[\text{M}/\text{H}]=-0.2$, as metallicity has a minimal impact on cluster CMDs and can be ignored; nevertheless, a value of $-0.2$ is still a fair average value for the Galactic anticentre \citep{Spina2022, 2024A&A...687A.239C}. In addition, as our considered clusters are too distant for their velocity dispersion to have an impact on their measured proper motion dispersion, we assume clusters are virialised (i.e. $\eta=10$) for this study.

Choosing how to explore this parameter space is non-trivial. On the one hand, we could define a fixed grid with eight co-ordinates ($l$, $b$, $d$, $M$, $\log t$, $r_c$, $\mu_{\alpha*}$, $\mu_\delta$), and perform cluster injection and retrievals on this grid. However, given the high dimensionality of this problem, defining such a grid is cumbersome, and it would be difficult to decide how to refine this grid later. In addition, this presents two further issues: firstly, we expect that the effects of parameters such as proper motion or cluster radius on a cluster's detectability should be broadly similar across the galaxy, meaning that a full grid-based approach could easily oversample certain parameters; and secondly, that whether or not a cluster is detected is somewhat stochastic, particularly given the probabilistic nature of the \cite{2019ApJ...887...93G} dust map that we sample extinctions from, meaning that a given grid co-ordinate has a distribution of values and would need to be sampled multiple times. Instead, we opted for a simpler approach for cluster simulations, and randomly sampled clusters from the parameter ranges defined in Table~\ref{tab:parameter_choices}, effectively performing our experiment on a randomly sampled unstructured grid. As long as clustering is performed enough times to fully sample the parameter space, this makes the process of cluster injection and retrieval easier, but it does require more careful post-processing of these results, which we discuss in Sect.~\ref{sec:cst_fdetected_predictor}.

We performed between 280 and 284 clustering runs on each of the 693 regions, each time injecting four randomly generated mock clusters into the \emph{Gaia} catalogue, which resulted in 194\,752 mock clusters in total. The clustering runs took a total of 6552 CPU hours on a CPU with a 3.7~GHz clock speed per core, or 205 hours ($\sim${}1~week) in total when ran on all 32 cores of the CPU simultaneously. Most of the computing time in this study was spent performing clustering analysis, which typically took around $\sim${}4 minutes per field; on the other hand, our pipeline only took $\sim${}1 second to generate each simulated cluster.

\subsection{Selection of optimum cluster detections}\label{sec:optimum_sel}

After performing cluster injection and retrievals, we selected the optimum \texttt{min\_cluster\_size} solution for each detected cluster, and cleaned our list of simulated detected clusters of certain poor-quality detections. The optimum \texttt{min\_cluster\_size} for each cluster was selected by finding the lowest \texttt{min\_cluster\_size} value that did not result in a cluster being split into multiple subclusters. In a select number of cases where the highest \texttt{min\_cluster\_size} value trialed in this work (80) still resulted in a cluster being split -- usually when outlying stars around a high-mass cluster were split into small, other subclusters -- only the detection with the highest CST was kept, with other duplicate clusters being dropped.

In total, 149\,914 of 194\,752 clusters were detected (i.e. have $\text{CST} \geq 3\sigma$). Of these detections, 1479 clusters have membership lists that include one or more stars from another simulated cluster in the same clustering run, as clustering was performed with four clusters per field per run. For now, we dropped these clusters from the final catalogue of simulated cluster retrievals as their detections may be biased, although a future study could investigate how overlap between clusters can impact their mutual detectability. In addition, 3173 clusters have members that overlap with real clusters in HR23. It is difficult to simulate whether or not a cluster in HR23 that actually contains multiple distinct clusters would have been correctly split into its constituent subclusters, as this was mostly a manual process: hence, we discarded 955 clusters with more than 50\% of their members belonging to a HR23 cluster, as the selection function when clusters strongly overlap is difficult to determine; the remaining 2218 clusters with overlap with HR23 clusters are retained and treated as positive cluster detections, as in this case, the simulated cluster (and not the HR23 cluster) would likely have been assigned as the main cluster in one of these pairs, and would hence have been detected.

Our final list of simulated clusters to analyse contains 192\,318 simulated clusters, 147\,639 of which were detected. This corresponds to around $\sim${}200\,clusters\,deg\textsuperscript{-2} across the area of the Galactic anticentre in this study, allowing us to study the cluster selection function at a resolution of up to $\sim${}1~degree.

\section{Results for the Galactic anticentre}\label{sec:results}

In this section, we summarise the results of our study, and investigate overall trends in our cluster injection and retrieval experiments, including which parameters make clusters more or less detectable. We use our injection and retrieval experiments to construct a cluster selection function model for the Galactic anticentre. Finally, we validate our results on real clusters.

\subsection{Overall trends in cluster detectability} \label{sec:results_trends}

\begin{figure*}
    \includegraphics[width=\textwidth]{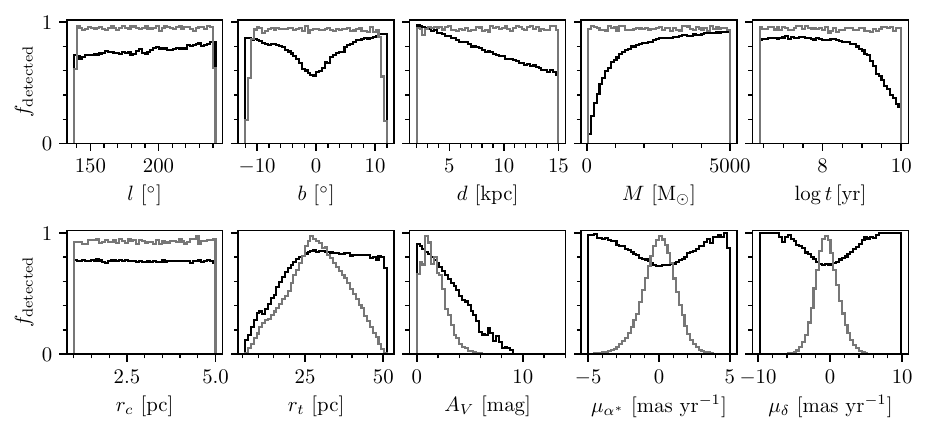}
    \caption{\label{fig:detection_trends_grid} Trends in cluster detectability as a function of ten cluster parameters for the simulated cluster injection and retrievals in this work. In each subplot, the black line shows the fraction of clusters detected, $f_\text{detected}$, while marginalising over all other parameters. For comparison, the grey line shows the distribution of injected clusters in this work, normalised to have a maximum at one. The top row of plots and the first on the lower row show intrinsic parameters, which are those chosen during this study; the remaining parameters $r_t$, $A_V$, $\mu_{\alpha*}$, and $\mu_\delta$ are all a function of these intrinsic parameters, but are nevertheless shown here for illustrative purposes. Figure~\ref{fig:detection_trends_cornerplot} shows the covariances between the most significant of these parameters.} 
\end{figure*}

\begin{figure}
    \includegraphics[width=\columnwidth]{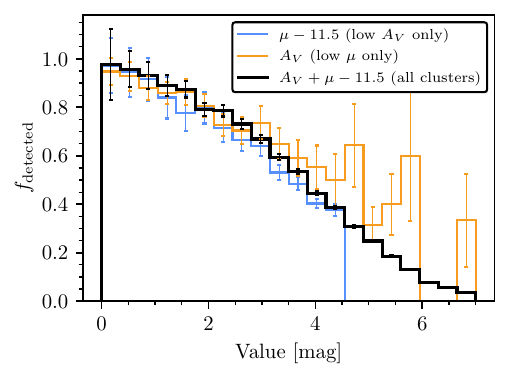}
    \caption{\label{fig:extinction_distance_comparison}Comparison between the impact of extinction $A_V$ and distance modulus $\mu$ on cluster detectability, for clusters with a mass below 800~M\textsubscript{\sun}. The blue curve shows $f_\text{detected}$ as a function of distance modulus $\mu$ for simulated clusters with negligible extinction ($A_V < 0.5$), while the orange curve shows $f_\text{detected}$ as a function of $A_V$ for the nearest clusters in this study ($2 < d \text{ [kpc]} < 3$). A correction of 11.5 (the distance modulus corresponding to a distance of 2~kpc) was subtracted from the distance modulus to make its value in this study start at zero, and hence be easily comparable with $A_V$, which also has a minimum value of zero in this work. The black curve shows $f_\text{detected}$ for $\mu$ and $A_V$ summed for all simulated clusters. Poisson uncertainties are shown on the bins.}
\end{figure}

\begin{figure*}
    \includegraphics[width=\textwidth]{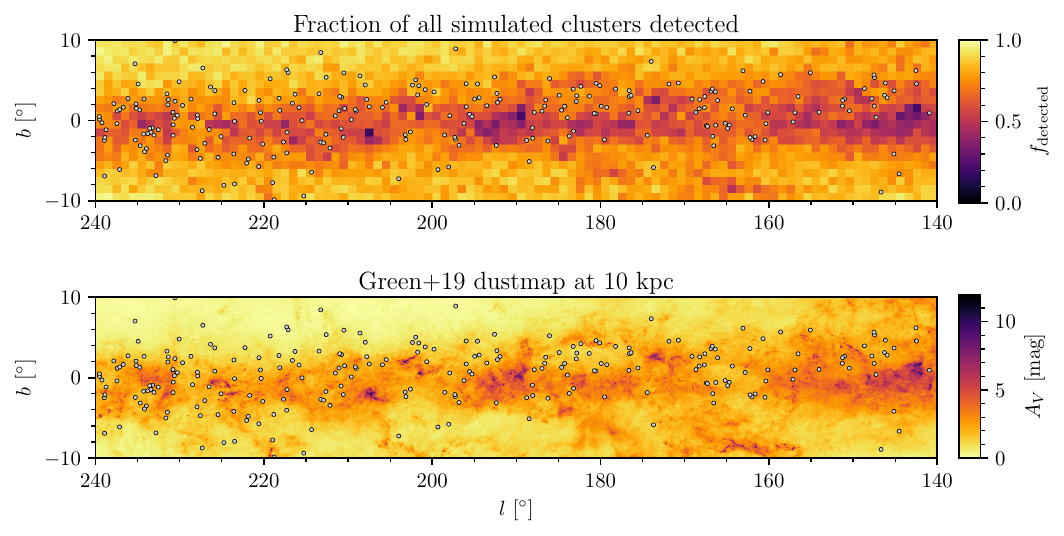}
    \caption{\label{fig:l_b_overall_detections} Fraction of simulated clusters recovered compared to the input dust distribution in this study. \emph{Top:} Fraction of all simulated clusters recovered in this work as a function of $l$ and $b$. Clusters were binned in bins of size $1^\circ\times1^\circ$. Grey circles show the distribution of high-quality OCs in HR24 with $CST>5$, $d>3$~kpc, and a CMD class of better than 0.5. \emph{Bottom:} As above, but instead showing the total \cite{2019ApJ...887...93G} extinction in the V band at a distance of 10~kpc overlaid with real OCs.}
\end{figure*}

\begin{figure*}
    \includegraphics[width=\textwidth]{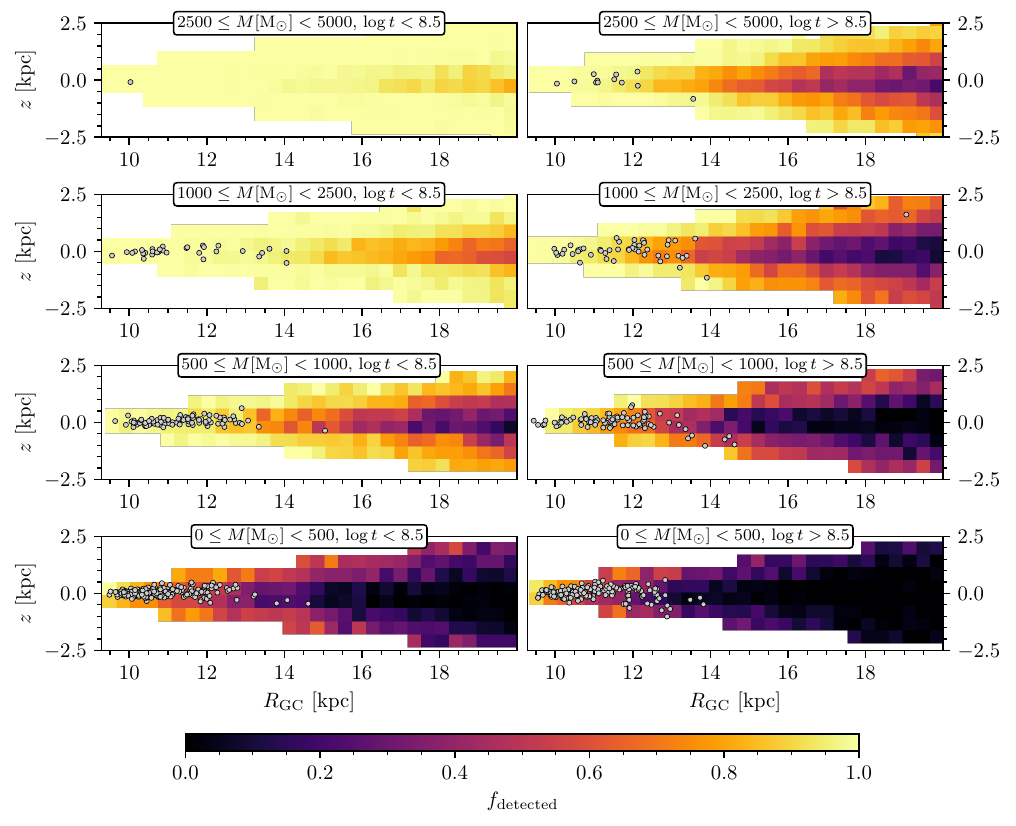}
    \caption{\label{fig:r_z_overall_detections}Fraction of simulated clusters recovered as a function of $R_\text{GC}$ and $Z$ divided into multiple different mass and ages ranges, and compared against the distribution of OCs in HR24 within those ranges. Each row shows clusters in a different mass range, indicated by the label on each subplot. The subplots in the left column show young clusters with $\log t < 8.5$, while subplots in the right column show old clusters with $\log t > 8.5$. Although proper motions only have a small impact on cluster detectability, we nevertheless only show the detection results of simulated clusters with proper motions $|\mu_{\alpha^*}| < 2.5$ and $|\mu_\delta| < 2.5$, providing a slightly more conservative estimate on cluster detectability at these locations.}
\end{figure*}

Simple binning of our simulated cluster detection and retrievals is able to reveal many details about cluster detectability in the Galactic anticentre, made possible as our cluster parameter ranges were uniformly sampled (Table~\ref{tab:parameter_choices}). Figure~\ref{fig:detection_trends_grid} shows the trends in cluster detectability as a function of ten key parameters. For further information, the covariances between the most significant parameters in Fig.~\ref{fig:detection_trends_grid} are shown in the corner plot of Fig.~\ref{fig:detection_trends_cornerplot}.

Cluster detectability strongly depends on cluster mass, as a higher-mass cluster will have more stars and be easier to see. The detectability increases logarithmically with mass, mirroring the result shown for the volume limited solar neighbourhood sample of HR24. Clusters become harder to detect as a linear function of distance, although very massive mock clusters are still detectable even at 15~kpc.

Extinction -- which is itself a function of $l$, $b$, and distance -- also has a strong effect on cluster detectability, owing to the strong impact that dust has on \emph{Gaia}'s visual-band observations. Clusters with $A_V \gtrsim 6$ appear difficult to detect in \emph{Gaia} DR3 data for the Galactic anticentre, with clusters with $A_V \gtrsim 9$ being impossible to detect even with the most favourable combination of all of their other parameters.

Cluster age presents a particularly interesting trend. Below ages of $\log t \approx 9$, which corresponds to 1~Gyr, cluster detectability does not depend significantly on age. However, cluster detectability drops off sharply above this age, with clusters at an age of 10~Gyr being more than twice as difficult to recover than clusters with an age below 1~Gyr. This is due to clusters containing progressively fewer bright stars as they age, reducing the number of sources visible above a given magnitude limit, which is likely to be a particular issue at the relatively high distances ($d \gtrsim 2$~kpc) considered in this study. The dependence on age also appears particularly covariant with cluster distance and extinction (Fig.~\ref{fig:detection_trends_cornerplot}). At a higher distance or extinction value, only stars beyond the turn-off point of an old cluster will be visible in \emph{Gaia}. Due to the power-law IMF and the fact that the time spent in the red-giant phase is $\lesssim 10\%$ of the main-sequence lifetime, the giant branch is much less populated than the main sequence. Thus, if only the giant stars of an old cluster are visible, its detectability is significantly lower.

Proper motions present a less strong but nevertheless important trend. Proper motions similar to that of the distant field stars in the anticentre ($\mu_{\alpha*} \approx \mu_\delta \approx 0$) somewhat impact cluster detectability, confirming that clusters on more extreme orbits are easier to detect because they stand out from the field stars in proper motion space.

The morphology of a cluster appears to have almost no impact on its detectability in our experiment. The strong trend for $r_t$ in Fig.~\ref{fig:detection_trends_grid} is due to the covariances between mass and distance on $r_t$, although this still suggests that structural parameter surveys of clusters not corrected for selection effects will under-report the prevalence of smaller (lower-mass) clusters. However, $r_c$ appears to have effectively no impact on the recoverability of clusters in HR23. This may be because proper motions and parallaxes are the most important factor for cluster recoverability with HDBSCAN, with the on-sky shape of a cluster being insignificant. In this study, we did not simulate the impact of crowding on the detectability of dense clusters; for OCs, this effect is negligible and is typically at the 1\% level \citep{2021A&A...649A...2L}. On the other hand, a study that includes clusters as dense and massive as globular clusters (which can be as much as $100 \times$ denser than the clusters in this study) may need to consider additional effects due to crowding on the \emph{Gaia} selection and astrometric accuracy functions, which would in turn impact cluster detectability \citep{2023A&A...669A..55C}. In this case, the core radius of clusters may begin to be at least slightly important, as it would affect how strongly crowded the cluster centre is. We also recall that the data mining strategy of HR23 (reproduced in this experiment) employs large tiles corresponding to HEALPix regions of level 5, which are roughly 2\,$^{\circ}$ wide. The apparent radius of a stellar overdensity might only play a role when it reaches a significant fraction of the investigated field of view. 

Extinction (measured in the V-band) and distance modulus appear to have a similar impact on cluster detectability with \emph{Gaia}. Figure~\ref{fig:extinction_distance_comparison} compares the impact of the distance modulus (adjusted to start at zero) and extinction of simulated clusters on the fraction of simulated clusters recovered in this study. For values below 3 mag, increases in $m-M$ or $A_V$ cause the same decreasing trend in cluster detectability, although highly extinguished clusters ($A_V \gtrsim 3$) seem easier to detect than clusters with a similarly high distance modulus. We explain the similar impact of extinction and distance modulus as being due to both parameters equally impacting how many stars are detectable with \emph{Gaia} for a given cluster, with increases to $m-M$ and $A_V$ causing a similar decrease in the number of visible stars due to \emph{Gaia}'s observations being in the optical. However, at higher distances, the impact of cluster distance on how well a cluster can be resolved in proper motions and parallaxes is a potential culprit for why detectability as a function of distance modulus drops off more steeply at values above 3 mag in Fig.~\ref{fig:extinction_distance_comparison}.

The on-sky distribution of recovered clusters shows clearly that clusters at low $|b|$ are harder to detect due to the impact of extinction in the disk. Figure~\ref{fig:l_b_overall_detections} shows a binned $f_\text{detected}$ map of all clusters in this study as a function of $l$ and $b$. This map clearly mirrors the trends in the \cite{2019ApJ...887...93G} dust map used to redden clusters in this work, with regions with high amounts of extinction corresponding to regions where clusters are harder to detect. In addition, the distribution of distant anticentre OCs in HR24 appears to have significant `holes' in regions where the \cite{2019ApJ...887...93G} dust map has high ($A_V \gtrsim 6$~mag) extinction, such as around $l\sim 190^\circ$ or $l\sim 142^\circ$. These regions correlate strongly with areas with a low $f_\text{detected}$.

Finally, the simple binned distribution of cluster recoveries as a function of $R_\text{GC}$ and $Z$ (shown in Fig.~\ref{fig:r_z_overall_detections}) shows how cluster detectability reduces as a function of $R_\text{GC}$, demonstrating the limits of OCs in \emph{Gaia} DR3 for tracing the Milky Way's cluster distribution. It is worth emphasising that our selection function depends on the heliocentric distance of a cluster from the Sun, and not on the cluster's galactocentric radius; nevertheless, the impact of increasing $R_\text{GC}$ is similar enough to heliocentric distance in the anticentre that this has only a minor impact on the figure. Figure~\ref{fig:r_z_overall_detections} shows that even in the highest mass range, old clusters are undetectable at high $R_\text{GC}$ and low $z$, suggesting that the old cluster census in the anticentre could be incomplete. We analyse this result further in Sect.~\ref{sec:discussion:scale_length} with a forward-modelling approach, and estimate the limiting radius that the Milky Way's OCs form at. We also comment further on the incompleteness of the old cluster population in Sect.~\ref{sec:discussion:berkeley_29_saurer_1}.

\subsection{Interpolating between simulated clusters on our unstructured grid of injection and retrievals}\label{sec:cst_fdetected_predictor}

Our cluster injection and retrievals densely sample the parameter space defined in Table~\ref{tab:parameter_choices}, allowing for an in-depth study of the cluster selection function towards the Galactic anticentre. However, it is difficult to use an unstructured probabilistic grid of detections to predict the detectability of a single cluster. It is possible to bin the unstructured grid to see broad overviews of detectability, such as binning in $l$, $b$, and $d$ and then `marginalising' over all other parameters to see detectability in 3D space; but it is also desirable to have a smooth selection function that allows for predictions at an exact set of cluster parameters. We explored a number of ways to interpolate between grid points on our unstructured grid of detections and construct a smooth selection function for other purposes in this study.

We initially explored the use of radial basis function and nearest neighbour interpolators. However, all interpolation schemes we tried were outperformed by supervised machine learning, with approaches based on the gradient-boosting decision tree algorithm XGBoost\footnote{\url{https://github.com/dmlc/xgboost}} \citep{xgboost_2016} outperforming the best interpolation scheme we trialed in terms of accuracy by a factor of $2.5$.

We trained multiple XGBoost predictors, tasked either with regression to predict CST values, or with classification to predict the probability that a cluster has a CST greater than some threshold. This effectively predicts the fraction of clusters with certain parameters detected at a given threshold, and can be written $f_{\text{detected},\text{CST} > x}$ for some threshold $x$. For $\text{CST}>3$, this recovers the selection function of HR23, who used such a threshold to clean their catalogue of poor-quality detections; higher thresholds are also useful to study the selection function of more astrometrically reliable clusters.

We first trained our models on nine parameters: ($l$, $b$, $d$, $M$, $\log t$, $r_c$, $r_t$, $\mu_{\alpha*}$, and $\mu_\delta$). However, since we found that $r_c$ and $r_t$ had a negligible effect on cluster detectability (see Fig.~\ref{fig:cst_shap}, and Sect. \ref{sec:results_trends} for further discussion), our final models do not incorporate them, and instead only require seven input parameters. 
For all trained XGBoost models, we used 160\,000 simulated clusters for training and kept the remaining 34\,752 clusters to use as a validation set. Our final models used 200 estimators, a maximum tree depth of seven, and used a learning rate (\texttt{eta}) of 0.2. Our CST predictor achieved final training and validation data root mean squared errors of 2.77 and 3.17 respectively (meaning that the typical errors of the predictor were around 2.77 and 3.17 respectively), with our $f_\text{detected}$ predictors achieving a typical binary classification accuracy on training and validation data of 94\% and 91\% respectively. Higher accuracies than this were not possible to achieve even with careful parameter tuning. Some of these residuals are due to sharp local variations in extinction and field density that XGBoost is not able to fully capture.

\subsection{Validating our results on real clusters}

\begin{figure}
    \includegraphics[width=\columnwidth]{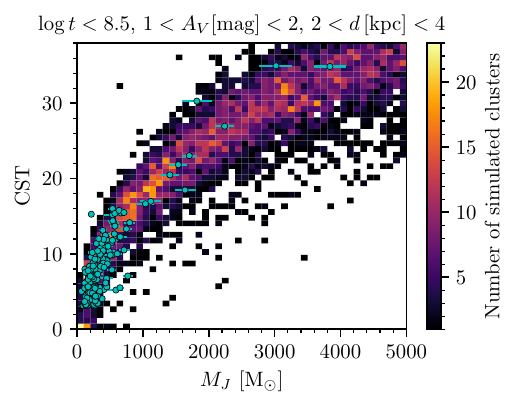}
    \caption{\label{fig:simulated_vs_real_detections} Comparison between the CST of simulated and real clusters in a restricted parameter range as a function of cluster mass. Simulated clusters are shown by the heatmap in the background, while clusters in HR23 are shown by the blue points, plotted as a function of their mass and associated uncertainty from HR24. We use the Jacobi mass $M_J$ from HR24, which measures the mass of the bound part of an OC, as $M_J$ is the most comparable cluster mass to the simulated cluster masses in this study. Only clusters with $\log t < 8.5$, $1 < A_V < 2$, and $2 < d < 4$~kpc are shown.}  
\end{figure}

We can now compare the CST predicted by our model to the values obtained by HR24 for clusters of similar parameters, providing an additional test on the reliability of our results.  Figure~\ref{fig:simulated_vs_real_detections} shows the distribution of cluster CSTs as a function of mass for a sample of young and moderately reddened anticentre clusters. The distribution of simulated cluster CSTs closely follows that of OCs in HR24, suggesting that our results are realistic.

For a more complete comparison, we used our CST predictor from Sect.~\ref{sec:cst_fdetected_predictor} to predict the CST of all real clusters in the anticentre. We found that our CST predictor -- trained only on our simulated cluster retrieval experiments -- had a root mean squared error of 3.75 when tasked with predicting the CST of real OCs in HR24. This is comparable to the root mean squared error of 3.17 that the CST predictor achieved on its simulated validation data, despite the fact that real OCs in HR24 have uncertain parameters.

\section{Answering questions about Galactic structure with our anticentre OC selection function}\label{sec:discussion}

In this section, we discuss the implications of the OC selection function on our understanding of the outer Galactic disk structure, traced through its cluster population. We comment on the different radial extent of the old and young OCs, on the warp of the disk, and on the potential future discovery of old clusters at very large Galactocentric distances.

\subsection{The limiting radius of the OC population as a function of age}\label{sec:discussion:scale_length}

\begin{figure}
    \includegraphics[width=\columnwidth]{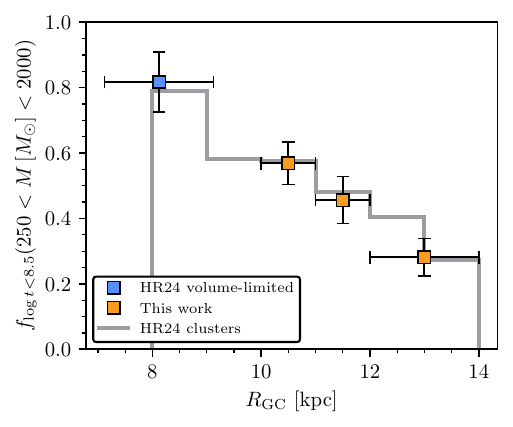}
    \caption{\label{fig:age_ratio} Fraction of OCs that are younger than $\log t=8.5$ as a function of galactocentric radius, shown only for clusters with masses between 250 and 2000~M\textsubscript{\sun}. Orange squares show this fraction estimated from cluster counts in HR24 and corrected for incompleteness assuming a `warm' kinematic heating model (Sect.~\ref{sec:discussion:scale_length}). The blue square is estimated for clusters within 1~kpc of the Sun in HR24, given that this volume is expected to be $\approx${}100\% complete within this mass range (see HR24 Sect.~5.1). The grey line shows this fraction for HR24 clusters in our studied region ($140^\circ< l < 240^\circ$, $|b| < 10^\circ$), uncorrected for selection effects.}
\end{figure}

The OC selection function we construct in this study indicates that at any given mass, beyond $\log t$$\sim$8.5, the detection probability quickly decreases with age. High proper motions and Galactic latitudes do not sufficiently increase detectability to balance the effect of old age. 

To quantify the effect of the selection function on an assumed cluster distribution, we generated two cluster populations, corresponding to two different heating regimes. The first one, which we refer to as the kinematically `warm' cluster population, follows a flared vertical density profile with a scale height increasing as $t^{1.3}$ \citep[][]{2020A&A...640A...1C}, and an age-velocity relation following \citet{2021A&A...647A..19T} which we add to the Milky Way rotation curve of the \texttt{MWPotential2014} potential of \citet{Bovy15}. The kinematically `hot' population scale height increases with $t^{2}$, and its velocity dispersion increases three times faster than in \citet{2021A&A...647A..19T}. The warm population is intended to represent a Milky-Way-like galaxy. The hot population is not meant as a physically plausible model of the Galactic disk, but as an illustration of the effect of extreme dynamical heating on cluster detectability. We convert their intrinsic properties to observable quantities (sky co-ordinates, proper motions) and apply the OC selection function to compute the fraction of recovered objects as a function of Galactocentric radius. 

Using our realistic `warm' model of OC kinematics in the Milky Way, Fig.~\ref{fig:age_ratio} shows the selection-effect corrected fraction of clusters younger than $\log t = 8.5$ as a function of galactocentric radius ($f_{\log t < 8.5}$). To create this figure, we binned clusters in HR24 with masses between 250 and 2000~M\textsubscript{\sun} into seven logarithmically spaced mass bins, three bins in distance, and two bins in age (`young' and `old'), and derived a distance, age, and mass-dependent correction for the count in each bin, averaged across the full anticentre region we consider. This correction was then used to scale the observed count of clusters in each bin to an estimated true count, based on the predicted detectability of $10^6$ clusters across this mass and distance range drawn from our `warm' kinematic model, before finally summing estimated true counts in all mass bins to derive the distribution as a function of age. An additional data point for the solar neighbourhood estimated using the volume-limited complete sample in HR24 is also shown. This result confirms that the observed higher fraction of old clusters in the anticentre is not a selection effect, but rather an intrinsic property of the Milky Way: we find that $f_{\log t < 8.5}$ is $2.97 \pm 0.11$ times lower at $R_\text{GC}=13$~kpc compared to in the solar neighbourhood, dropping from $81.7\%\pm9.2\%$ in the solar neighbourhood to $27.5\%\pm5.7\%$ in this most distant bin.

Curiously, our selection effect-corrected $f_{\log t < 8.5}$ remains relatively similar to the observed distribution of this quantity in HR24 -- at least within our considered distance and mass ranges. Only the highest distance bin in Fig.~\ref{fig:age_ratio} appears significantly different to the observed distribution (i.e. the grey line), likely because it corresponds to the distance range in which old clusters start to become significantly more challenging to recover compared to young clusters (Fig.~\ref{fig:r_z_overall_detections}). Cluster mass appears to remain the most significant parameter impacting cluster detectability, and not age (see also Appendix~\ref{app:interpretation}), causing the simple ratio between young and old clusters within some distant range to not be strongly biased by detection effects due to age. Estimating $f_{\log t < 8.5}$ at distances greater than 14~kpc is not possible with our simple binning-based approach due to the small number statistics of young clusters at this distance; a forward modelling-based approach should be explored in the future using our selection function to allow for a smooth model to be fit to $f_{\log t < 8.5}$.

\begin{figure*}[t]
    \centering
    \includegraphics[width=0.99\textwidth]{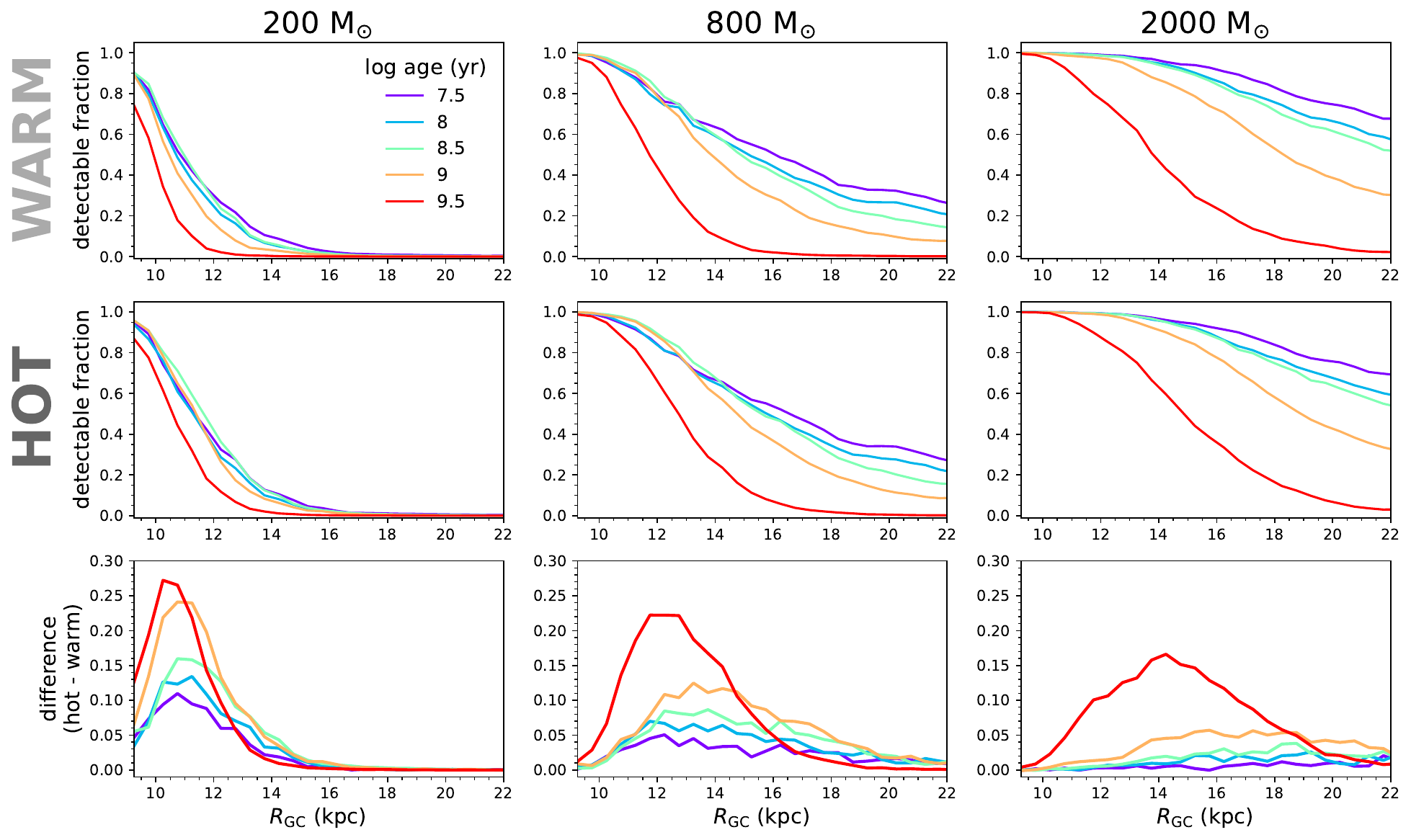} 
    \caption{\label{fig:frac_Rgc_threemasses} \emph{Top:} Fraction of detectable clusters as a function of Galactocentric radius at three different masses and five different ages, for the simulated kinematically warm cluster population (see text). \emph{Middle:} Same as above, for the kinematically hot cluster population. \emph{Bottom:} Difference between hot and warm detectability fractions.
    }
\end{figure*}

Figure~\ref{fig:age_ratio} also shows that the greater scale height and hotter orbits of old clusters are not significant enough to explain their higher fraction in the anticentre. To further illustrate the implausibility of $f_{\log t < 8.5}$ being flat as a function of $R_\text{GC}$, Fig.~\ref{fig:frac_Rgc_threemasses} shows the impact of assuming our unrealistically `hot' kinematics on the detectability of OCs for three chosen masses (200, 800, and 2000\,$M_{\odot}$). While the hot cluster population allows for a higher detection fraction at a given age than the warm population, its extreme combination of large scale heights and hot proper motions is never sufficient to make old clusters significantly more detectable than younger ones. Based on our current work, it seems unlikely that the larger relative number of known old clusters in the outer Milky Way (Fig.~\ref{fig:XY_RgcZ}) is caused by an observational bias favouring old clusters, with it instead being more likely that young clusters are either scarcer than old clusters or have lower masses. Other age-dependent effects could be investigated in the future (such as mass segregation or alternative cluster models to a King model), although they would have to make up for a large difference in cluster detectability as a function of age to counter the result in this work.

The diminishing number of young clusters with increasing Galactocentric radius is however not an indication of a lack of star formation, as young field stars are visible as far as 18\,kpc from the Galactic centre \citep[e.g. Cepheids,][]{2022A&A...668A..40L}. We therefore argue that R$_{\mathrm{GC}}$$\sim$13\,kpc could be the limit for massive cluster formation, rather than the absolute cut-off radius of star formation in the Milky Way. While our experiment confidently establishes that old massive clusters are more numerous than young massive clusters in the outer disc, it also demonstrates that the \textit{Gaia}-DR3-based cluster census is highly incomplete for distant low-mass clusters of all ages (bottom panels of Fig.~\ref{fig:r_z_overall_detections}; top left panel of Fig.~\ref{fig:frac_Rgc_threemasses}). 

Rather than a constraint on the total number of young low-mass clusters, our results \textbf{could} be interpreted as an upper limit on the mass of clusters forming in the outer Milky Way. This interpretation supports the theory proposed by \citet{2008Natur.455..641P} to explain why the radial density profiles of external galaxies exhibit a sharp edge in H$\alpha$, but a much smoother UV (ultraviolet) profile. The authors propose that lower gas densities reduce the maximum mass that clusters can form at. 
In this scenario, the massive old clusters currently found in the outer-disc formed in the inner Milky Way before migrating outwards, where weaker tidal forces \citep{Lamers2005A&A...441..117L} and the larger vertical excursion \citep{Viscasillas2023A&A...679A.122V,Moreira2025A&A...694A..70M} increased the timescale for their disruption.

Migration would also be responsible for the flattening of the radial metallicity gradient, whose change of slope occurs near R$_{\mathrm{GC}}$$\sim$12\,kpc \citep{Donor2020,Zhang2021ApJ...919...52Z,Myers2022AJ....164...85M,Spina2022,Magrini2023,2024A&A...687A.239C}. Studying migration mechanisms in the Milky Way through the lens of OC distributions is complex, as mild perturbations to their orbits may enhance their chances of survival, but strong perturbations can accelerate their disruption. Recent studies estimate the migration rates of OCs around 1\,kpc\,Gyr$^{-1}$ \citep{ChenZhao2020MNRAS.495.2673C,Netopil2022MNRAS.509..421N}. \citet{ChenZhao2020MNRAS.495.2673C} remark that the metallicities and kinematics of old outer-disc clusters suggest higher migration rates of 1.5$\pm$0.5\,kpc\,Gyr$^{-1}$, further supporting the idea that clusters undergoing significant outward migration are more likely to remain bound for billions of years.

However, caution should be taken when dealing with the small sample size of young, massive clusters in the anticentre due to the rarity of young, massive clusters \citep{lada_embedded_2003, 2010ARAA..48..431P}. Our result mirrors discussions in extragalactic star cluster samples: while some works find a galactocentric-radius dependent limiting mass for clusters \cite[e.g.][]{2013MNRAS.435.2604P}, other works have demonstrated that a small sample size can cause the appearance of a truncated upper mass limit for clusters \citep{2016ApJ...816....9S}. In general, robustly determining the upper mass limit of cluster censuses is challenging \citep{KrumholzMcKee_2019}. While we are able to conclude in this work that the anticentre contains more old clusters than young ones, further research will be required to determine the exact reason for this. Alternative hypotheses to cluster migration should be explored, such as a lower cluster disruption rate in the anticentre causing an `old-heavy' cluster age function relative to the solar neighbourhood.

\subsection{The influence of the Galactic warp on the distribution of clusters}\label{sec:discussion:warp}
 
The Milky Way features a warped distribution of neutral hydrogen gas in the anticentre \citep{2007A&A...469..511K,2006ApJ...643..881L}, which has also been shown with the distribution of Cepheid variable stars \citep[e.g.][]{2014Natur.509..342F,Skowron2019Sci...365..478S,2019NatAs...3..320C,2022A&A...668A..40L}, young OCs \citep[e.g.][]{MoitinhoVazquez_2006,VazquezMay_2008}, old OCs \citep[e.g.][]{2020A&A...640A...1C,He2023warp}, and red clump stars \citep[e.g.][]{MomanyZaggia_2006}. The warp is also visible in the old cluster distribution in Fig.~\ref{fig:XY_RgcZ}. In HR24 and using ages from \cite{2024AJ....167...12C}, restricting analysis to old ($\log t > 8.8$) OCs with $12 < R_\text{GC} \; \text{[kpc]} < 16$, and using a slightly higher CST cut to improve cluster reliability ($\text{CST} > 4\sigma$), there are just six clusters above the disk ($500 > Z \text{\;[pc]} > 2000$) compared to 21 below the disk ($-500 < Z \text{\;[pc]} < -2000$). This difference in cluster counts can be interpreted as being due to the Galactic warp; however, if a selection bias towards the anticentre existed that made clusters, stars, and gas harder to detect above the disk than below the disk, flaring of the disk with an obscured upper part would present as an asymmetric warp.

As a simple test of the selection dependence of the observed warped cluster distribution, we tested whether it is easier to detect the population of below-disk clusters below or above the disk. To do so, we calculated the probability of detecting the below-disk population of clusters at $\text{CST} > 4\sigma$ below and above the disk by taking the 21 below disk clusters and simply flipping the sign of their $Z$ co-ordinate, and then calculating the mean $f_{\text{detected},\text{CST}>4}$ value of these clusters with our model from Sect.~\ref{sec:cst_fdetected_predictor}, in addition to comparing this to their $f_{\text{detected},\text{CST}>4}$ below the disk. The mean detection probability of the below disk clusters is 0.40, suggesting that most of the below-disk cluster population in HR24 was somewhat difficult to detect; on the other hand, it would be nearly twice as easy to see the below-disk cluster population above the disk, as the mean detection probability of the below-disk clusters if they were instead above the disk would be 0.74. This difference is likely due to the generally higher extinction below the disk than above it in the direction of the Galactic anticentre (see lower panel of Fig.~\ref{fig:l_b_overall_detections}.)

Correcting for selection effects, we would expect to see $52.5 \pm 11.4$ clusters below the disk and $8.1 \pm 3.3$ clusters above the disk with similar parameters to the 21 below the disk in HR24. Adding the uncertainties on these quantities together in quadrature, we find a statistical difference between these counts at the 3.7$\sigma$ level, which is strong evidence that the warp in the OC census is not a selection effect. In fact, if all clusters were visible, then the warp in the OC census would present itself nearly twice as strongly as it currently does in the old OC distribution.

\subsection{The isolation of Saurer 1 and Berkeley 29}\label{sec:discussion:berkeley_29_saurer_1}

\begin{figure}
    \includegraphics[width=\columnwidth]{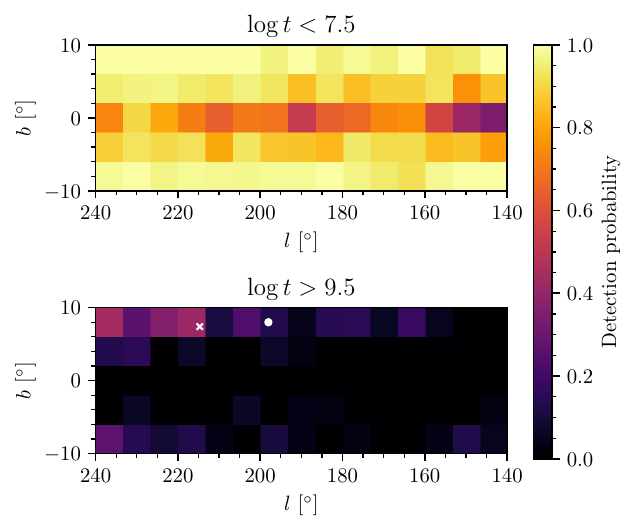}
    \caption{\label{fig:l_b_berkeley_29_saurer_1} Fraction of simulated clusters recovered at young and old ages compared to the location of old distant clusters Berkeley~29 and Saurer~1. \emph{Top:} Sky distribution of simulated cluster recoveries for young clusters $\log t < 7.5$, at distances greater than 10~kpc, and with masses in the range $1500 < M < 2500$~M\textsubscript{\sun}. \emph{Bottom:} As above, but for old clusters with $\log t > 9.5$. The location of Berkeley~29 is indicated by a white circle; the location of Saurer~1 is indicated by a white cross.} 
\end{figure}

Our cluster selection function also allows for the study of OCs at a more individual level. Arguably the most puzzling objects in the Galactic anticentre are Berkeley~29 and Saurer~1, two old ($\log t > 9.5$) OCs at a significantly larger $R_\text{GC}$ than any other known OC (see Fig.~\ref{fig:XY_RgcZ}). These objects are just about visible in \emph{Gaia} DR3 \citep{GaiaEDR3Anticentre,PerrenDistantOCs2022}, although only Berkeley~29 appears in HR23 and HR24 and with a modest CST of 5.99. In \cite{GaiaEDR3Anticentre}, it appears that Saurer~1 has fewer stars than Berkeley~29, suggesting that it is smaller and thus harder to detect due to its lower mass. 

Figure~\ref{fig:r_z_overall_detections} shows that the detection of Berkeley~29 by HR23 may be somewhat lucky and depend on its position in the galaxy. Old clusters are almost impossible to detect at low $Z$ values at the distance of Berkeley~29, particularly if their mass is below 1000~M\textsubscript{$\sun$}. Its location at a high $b$ of 7.98$^\circ$ above the dusty plane of the Milky Way may be the only reason why it is visible. Between where the main OC distribution truncates at $R_{GC} \sim 14$~to~15~kpc and the 19~kpc distance of Berkeley~29, it is easy to imagine that many old clusters could hide in the dust-obscured void in Fig.~\ref{fig:r_z_overall_detections} at low $|Z|$. In the solar neighbourhood, old clusters with masses of just a few hundred solar masses are not especially rare (such as the Hyades or Ruprecht~147), with low-mass old clusters in fact being more common than high-mass ones (HR24, Fig. 15); a similar mass distribution in the anticentre would imply dozens of low-mass or low-$|Z|$ clusters are waiting to be discovered.

The detection of Berkeley~29 may also be dependent on its location in Galactic longitude along the disk. Figure~\ref{fig:l_b_berkeley_29_saurer_1} shows the detectability of old and relatively high-mass clusters with parameters similar to Berkeley~29. As with the Galactic warp in Sect.~\ref{sec:discussion:warp}, it is around twice as easy to detect Berkeley~29 above the disk than below it, largely due to extinction -- suggesting that Berkeley~29 could have many low-$Z$ counterparts. In addition, the left side of the anticentre ($l > 180^\circ$) appears to be the easiest place to see distant old clusters in, with the upper-right of the anticentre being significantly more difficult than the upper-left. In 3D, it appears that Saurer~1 and Berkeley~29 are in the easiest part of the distant anticentre to detect clusters, and we hypothesise that deeper astrometric and photometric surveys of the anticentre such as {\it Gaia} DR4 or the proposed GaiaNIR mission \citep{GaiaNIR2018} will find additional clusters in this region. The southern hemisphere location of the Vera Rubin Observatory is better suited to observe the Galactic centre, but LSST may provide enough deep astrometric and photometric data in the medium term of some anticentre regions to allow for the discovery of more anticentre clusters \citep{2019ApJ...873..111I,2023PASP..135g4201U}, particularly when using co-added images that could reach up to six magnitudes deeper than \emph{Gaia}'s $G \approx 20.7$ magnitude limit.

No other OCs as distant as these are known in the Galactic anticentre, which poses the question of whether Berkeley~29 and Saurer~1 are exceptions to the OC distribution, or whether there is a population of distant and old relic OCs in the anticentre yet to be discovered due to selection effects. Improving our knowledge of the old residents of the outer-disc would help us put better constraints on the efficiency of the mechanisms driving stellar migration, and their role in preserving bound clusters. Once believed to be of extragalactic origin, all currently known old distant clusters (including Berkeley~29 and Saurer~1) have been shown beyond reasonable doubt to have formed inside the Milky Way \citep[e.g.][]{2007A&A...476..217C,2016A&A...588A.120C,GaiaEDR3Anticentre}. The present study also shows that these two clusters, once thought to be `alien', might very well be the tip of the iceberg of a population of distant outer galaxy clusters that are as yet undiscovered.

\section{Conclusions and future outlook} \label{sec:conclusion}

In this paper, we demonstrated injection and retrieval of simulated clusters as a way to determine an empirical star cluster selection function for the Milky Way, beginning with the present study focusing on the Galactic anticentre. We began by generating a large sample of mock clusters spanning a wide range of parameters, and injected them into the \emph{Gaia}~DR3 catalogue in the direction of the Galactic anticentre. We then ran the HDBSCAN-based OC detection strategy of HR23 and attempted to recover the simulated clusters. We built a model of detection probabilities as a function of the main cluster parameters, thus constructing the cluster selection function of the HR24 catalogue in the Galactic anticentre. 

We investigated the impact of different cluster parameters, finding that the mass, distance, extinction, and age of a cluster are the most fundamental parameters influencing its detectability. In addition, the proper motion of a cluster can also have a sizeable impact on its detectability, with clusters with the most distinct velocity from that of the field stars around it being easier to detect. We also show that, for \emph{Gaia}'s broad-band optical observations, distance modulus and extinction have a similar level of impact on cluster detectability. In this work, we do not find that cluster detectability depends critically on the radial profile of a cluster, although this should be investigated further in future works, including exploring the impact of effects such as mass segregation.

Using our cluster selection function, we show that old clusters are significantly harder to recover than young clusters in the outer Galaxy, despite catalogues containing a larger number of old clusters in the outer-disc. We conducted forward-modelling of the cluster incompleteness and find that clusters older than $\log t=8.5$ are $2.97\pm0.11$ times more common at a Galactocentric radius of 13~kpc than in the solar neighbourhood -- despite other positive detection biases in their favour, such as hotter orbits or a higher scale height. We conclude from this experiment that few young clusters remain to be found at large Galactocentric distances, while the old cluster census is potentially very incomplete. In particular, we show that the most distant known clusters in the Galactic anticentre -- Berkeley~29 and Saurer~1 -- are likely to have been detected because of their favourably high Galactic latitude away from the dusty plane of the Milky Way. We expect that these two clusters may be the tip of the iceberg of a population of distant, old clusters in the Galactic anticentre. We interpret the presence of a population of old massive clusters in the distant outer regions of the Milky Way as a combination of efficient migration mechanisms and a longer cluster survival time in the outer-disc.

Our work is not without some caveats that future studies should address. Firstly, our work implicitly assumes that object classifications in HR24 are correct. Principally, this requires that cluster masses and radii in HR24 are correct enough to avoid misclassifying bound clusters as unbound ones. While this has a small impact on our studied distant region in the anticentre (Table~\ref{tab:catalogues}), this will need to be addressed further when extending our methodology to regions of HR24 that contain a higher number of moving groups. Secondly, our work does not establish the selection function of the HR23 CMD classifier, and hence cannot yet be used to determine the selection function of their high-quality cluster sample. Finally, since our selection function depends strongly on a number of observables that are difficult to calculate accurately for clusters (such as mass, age, or extinction), the applicability of our selection function is also somewhat limited by the quality of cluster parameters. Efforts should be made to continue improving cluster parameter determination in the future, particularly by using upcoming data releases from \emph{Gaia} and other surveys. In addition, better quantifying the systematics of these parameter determinations would allow for corrections for wrongly assigned parameters to be incorporated into future cluster selection functions.

The methodology we developed for this study can be applied to the entire \emph{Gaia} catalogue. In a future work, we will address the computational challenges involved in applying our method across the whole \emph{Gaia} catalogue, and we will release an open-source selection function of the HR24 OC catalogue. A key application of this selection function will be in constructing a full model of the currently unknown age-mass distribution of OCs across the full Milky Way, allowing for OCs to trace cluster formation and destruction rates across the full Galactic disk. In addition to increasing the utility of OCs in mapping Galactic structure, our cluster selection function products and methodology could be used for many other purposes, such as: determining how many stars are missing from a given cluster, calculating the detection probability of OC tidal tails, or estimating how many clusters future surveys or data releases may be able to detect. We plan to explore these possibilities in future works.

\begin{acknowledgements}

We thank the anonymous referee for their detailed comments and feedback which greatly improved the clarity of this work.
We also thank the organizers of the ``From star clusters to field populations: survived, destroyed and migrated clusters'' workshop in 2023 in Florence, Italy, which enabled discussions and collaborations that led to this paper.
This work has made use of data from the European Space Agency (ESA) mission {\it Gaia} (\url{http://www.cosmos.esa.int/gaia}), processed by the {\it Gaia} Data Processing and Analysis Consortium (DPAC,
\url{http://www.cosmos.esa.int/web/gaia/dpac/consortium}). Funding for the DPAC has been provided by national institutions, in particular the institutions participating in the {\it Gaia} Multilateral Agreement. 
This work is a result of the GaiaUnlimited project, which has received funding from the European Union’s Horizon 2020 research and innovation programme under grant agreement No 101004110. The GaiaUnlimited project was started at the 2019 Santa Barbara \emph{Gaia} Sprint, hosted by the Kavli Institute for Theoretical Physics at the University of California, Santa Barbara. 
This work was (partially) supported by the Spanish MICIN/AEI/10.13039/501100011033 and by "ERDF A way of making Europe" by the European Union through grant PID2021-122842OB-C21, and the Institute of Cosmos Sciences University of Barcelona (ICCUB, Unidad de Excelencia ’Mar\'{\i}a de Maeztu’) through grant CEX2019-000918-M. FA acknowledges the grant RYC2021-031683-I funded by MCIN/AEI/10.13039/501100011033 and by the European Union's NextGenerationEU/PRTR.
In addition to those listed in the text, this work made use of the following software packages: \texttt{Jupyter} \citep{2007CSE.....9c..21P, kluyver2016jupyter}, \texttt{matplotlib} \citep{Hunter:2007}, \texttt{numpy} \citep{numpy}, \texttt{pandas} \citep{mckinney-proc-scipy-2010, pandas_10957263}, \texttt{python} \citep{python}, \texttt{scipy} \citep{2020SciPy-NMeth, scipy_10909890}, \texttt{astroquery} \citep{2019AJ....157...98G, astroquery_10799414}, \texttt{Cython} \citep{cython:2011}, \texttt{Numba} \citep{numba:2015, Numba_10839385}, and \texttt{scikit-learn} \citep{scikit-learn, sklearn_api, scikit-learn_10951361}. This research has made use of NASA's Astrophysics Data System. This research also made use of the SIMBAD database, operated at CDS, Strasbourg, France \citep{wengerm_simbad_astronomical_2000}. Software citation information aggregated using \texttt{\href{https://www.tomwagg.com/software-citation-station/}{The Software Citation Station}} \citep{software-citation-station-paper, software-citation-station-zenodo}. This work used the accessible \texttt{Matplotlib}-like colour cycles defined in \cite{petroff_accessible_color_2021}.

\end{acknowledgements}

\bibliographystyle{aa}
\bibliography{refs}

\begin{thebibliography}{112}
\expandafter\ifx\csname natexlab\endcsname\relax\def\natexlab#1{#1}\fi

\bibitem[{{Alessi} {et~al.}(2003){Alessi}, {Moitinho}, \&
  {Dias}}]{2003A&A...410..565A}
{Alessi}, B.~S., {Moitinho}, A., \& {Dias}, W.~S. 2003, \aap, 410, 565

\bibitem[{{Almeida} {et~al.}(2023){Almeida}, {Monteiro}, \&
  {Dias}}]{AlmeidaMonteiro_2023}
{Almeida}, A., {Monteiro}, H., \& {Dias}, W.~S. 2023, Monthly Notices of the
  Royal Astronomical Society, 525, 2315

\bibitem[{{Almeida} {et~al.}(2025){Almeida}, {Moitinho}, \&
  {Moreira}}]{AlmeidaMoitinho_2025}
{Almeida}, D., {Moitinho}, A., \& {Moreira}, S. 2025, Astronomy and
  Astrophysics, 693, A305

\bibitem[{{Anders} {et~al.}(2021){Anders}, {Cantat-Gaudin}, {Quadrino-Lodoso},
  {Gieles}, {Jordi}, {Castro-Ginard}, \&
  {Balaguer-N{\'u}{\~n}ez}}]{2021A&A...645L...2A}
{Anders}, F., {Cantat-Gaudin}, T., {Quadrino-Lodoso}, I., {et~al.} 2021, \aap,
  645, L2

\bibitem[{{Astropy Collaboration} {et~al.}(2022){Astropy Collaboration},
  {Price-Whelan}, {Lim}, {Earl}, {Starkman}, {Bradley}, {Shupe}, {Patil},
  {Corrales}, {Brasseur}, {N{"o}the}, {Donath}, {Tollerud}, {Morris},
  {Ginsburg}, {Vaher}, {Weaver}, {Tocknell}, {Jamieson}, {van Kerkwijk},
  {Robitaille}, {Merry}, {Bachetti}, {G{"u}nther}, {Aldcroft},
  {Alvarado-Montes}, {Archibald}, {B{'o}di}, {Bapat}, {Barentsen}, {Baz{'a}n},
  {Biswas}, {Boquien}, {Burke}, {Cara}, {Cara}, {Conroy}, {Conseil}, {Craig},
  {Cross}, {Cruz}, {D'Eugenio}, {Dencheva}, {Devillepoix}, {Dietrich},
  {Eigenbrot}, {Erben}, {Ferreira}, {Foreman-Mackey}, {Fox}, {Freij}, {Garg},
  {Geda}, {Glattly}, {Gondhalekar}, {Gordon}, {Grant}, {Greenfield}, {Groener},
  {Guest}, {Gurovich}, {Handberg}, {Hart}, {Hatfield-Dodds}, {Homeier},
  {Hosseinzadeh}, {Jenness}, {Jones}, {Joseph}, {Kalmbach}, {Karamehmetoglu},
  {Ka{l}uszy{'n}ski}, {Kelley}, {Kern}, {Kerzendorf}, {Koch}, {Kulumani},
  {Lee}, {Ly}, {Ma}, {MacBride}, {Maljaars}, {Muna}, {Murphy}, {Norman},
  {O'Steen}, {Oman}, {Pacifici}, {Pascual}, {Pascual-Granado}, {Patil},
  {Perren}, {Pickering}, {Rastogi}, {Roulston}, {Ryan}, {Rykoff}, {Sabater},
  {Sakurikar}, {Salgado}, {Sanghi}, {Saunders}, {Savchenko}, {Schwardt},
  {Seifert-Eckert}, {Shih}, {Jain}, {Shukla}, {Sick}, {Simpson},
  {Singanamalla}, {Singer}, {Singhal}, {Sinha}, {Sip{H{o}}cz}, {Spitler},
  {Stansby}, {Streicher}, {{{S}}umak}, {Swinbank}, {Taranu}, {Tewary},
  {Tremblay}, {Val-Borro}, {Van Kooten}, {Vasovi{'c}}, {Verma}, {de Miranda
  Cardoso}, {Williams}, {Wilson}, {Winkel}, {Wood-Vasey}, {Xue}, {Yoachim},
  {Zhang}, {Zonca}, \& {Astropy Project Contributors}}]{astropy:2022}
{Astropy Collaboration}, {Price-Whelan}, A.~M., {Lim}, P.~L., {et~al.} 2022,
  \apj, 935, 167

\bibitem[{{Astropy Collaboration} {et~al.}(2018){Astropy Collaboration},
  {Price-Whelan}, {Sip{\H{o}}cz}, {G{\"u}nther}, {Lim}, {Crawford}, {Conseil},
  {Shupe}, {Craig}, {Dencheva}, {Ginsburg}, {Vand erPlas}, {Bradley},
  {P{\'e}rez-Su{\'a}rez}, {de Val-Borro}, {Aldcroft}, {Cruz}, {Robitaille},
  {Tollerud}, {Ardelean}, {Babej}, {Bach}, {Bachetti}, {Bakanov}, {Bamford},
  {Barentsen}, {Barmby}, {Baumbach}, {Berry}, {Biscani}, {Boquien}, {Bostroem},
  {Bouma}, {Brammer}, {Bray}, {Breytenbach}, {Buddelmeijer}, {Burke},
  {Calderone}, {Cano Rodr{\'\i}guez}, {Cara}, {Cardoso}, {Cheedella}, {Copin},
  {Corrales}, {Crichton}, {D'Avella}, {Deil}, {Depagne}, {Dietrich}, {Donath},
  {Droettboom}, {Earl}, {Erben}, {Fabbro}, {Ferreira}, {Finethy}, {Fox},
  {Garrison}, {Gibbons}, {Goldstein}, {Gommers}, {Greco}, {Greenfield},
  {Groener}, {Grollier}, {Hagen}, {Hirst}, {Homeier}, {Horton}, {Hosseinzadeh},
  {Hu}, {Hunkeler}, {Ivezi{\'c}}, {Jain}, {Jenness}, {Kanarek}, {Kendrew},
  {Kern}, {Kerzendorf}, {Khvalko}, {King}, {Kirkby}, {Kulkarni}, {Kumar},
  {Lee}, {Lenz}, {Littlefair}, {Ma}, {Macleod}, {Mastropietro}, {McCully},
  {Montagnac}, {Morris}, {Mueller}, {Mumford}, {Muna}, {Murphy}, {Nelson},
  {Nguyen}, {Ninan}, {N{\"o}the}, {Ogaz}, {Oh}, {Parejko}, {Parley}, {Pascual},
  {Patil}, {Patil}, {Plunkett}, {Prochaska}, {Rastogi}, {Reddy Janga},
  {Sabater}, {Sakurikar}, {Seifert}, {Sherbert}, {Sherwood-Taylor}, {Shih},
  {Sick}, {Silbiger}, {Singanamalla}, {Singer}, {Sladen}, {Sooley},
  {Sornarajah}, {Streicher}, {Teuben}, {Thomas}, {Tremblay}, {Turner},
  {Terr{\'o}n}, {van Kerkwijk}, {de la Vega}, {Watkins}, {Weaver}, {Whitmore},
  {Woillez}, {Zabalza}, \& {Astropy Contributors}}]{astropy:2018}
{Astropy Collaboration}, {Price-Whelan}, A.~M., {Sip{\H{o}}cz}, B.~M., {et~al.}
  2018, \aj, 156, 123

\bibitem[{{Astropy Collaboration} {et~al.}(2013){Astropy Collaboration},
  {Robitaille}, {Tollerud}, {Greenfield}, {Droettboom}, {Bray}, {Aldcroft},
  {Davis}, {Ginsburg}, {Price-Whelan}, {Kerzendorf}, {Conley}, {Crighton},
  {Barbary}, {Muna}, {Ferguson}, {Grollier}, {Parikh}, {Nair}, {Unther},
  {Deil}, {Woillez}, {Conseil}, {Kramer}, {Turner}, {Singer}, {Fox}, {Weaver},
  {Zabalza}, {Edwards}, {Azalee Bostroem}, {Burke}, {Casey}, {Crawford},
  {Dencheva}, {Ely}, {Jenness}, {Labrie}, {Lim}, {Pierfederici}, {Pontzen},
  {Ptak}, {Refsdal}, {Servillat}, \& {Streicher}}]{astropy:2013}
{Astropy Collaboration}, {Robitaille}, T.~P., {Tollerud}, E.~J., {et~al.} 2013,
  \aap, 558, A33

\bibitem[{Behnel {et~al.}(2011)Behnel, Bradshaw, Citro, Dalcin, Seljebotn, \&
  Smith}]{cython:2011}
Behnel, S., Bradshaw, R., Citro, C., {et~al.} 2011, Computing in Science
  Engineering, 13, 31

\bibitem[{{Bovy}(2015)}]{Bovy15}
{Bovy}, J. 2015, \apjs, 216, 29

\bibitem[{{Bressan} {et~al.}(2012){Bressan}, {Marigo}, {Girardi}, {Salasnich},
  {Dal Cero}, {Rubele}, \& {Nanni}}]{Bressan2012MNRAS.427..127B}
{Bressan}, A., {Marigo}, P., {Girardi}, L., {et~al.} 2012, \mnras, 427, 127

\bibitem[{{Brown}(2021)}]{2021ARA&A..59...59B}
{Brown}, A. G.~A. 2021, \araa, 59, 59

\bibitem[{Buitinck {et~al.}(2013)Buitinck, Louppe, Blondel, Pedregosa, Mueller,
  Grisel, Niculae, Prettenhofer, Gramfort, Grobler, Layton, VanderPlas, Joly,
  Holt, \& Varoquaux}]{sklearn_api}
Buitinck, L., Louppe, G., Blondel, M., {et~al.} 2013, in ECML PKDD Workshop:
  Languages for Data Mining and Machine Learning, 108--122

\bibitem[{Campello {et~al.}(2013)Campello, Moulavi, \& Sander}]{hdbscan_paper}
Campello, R. J. G.~B., Moulavi, D., \& Sander, J. 2013, in Advances in
  Knowledge Discovery and Data Mining, ed. J.~Pei, V.~S. Tseng, L.~Cao,
  H.~Motoda, \& G.~Xu (Berlin, Heidelberg: Springer Berlin Heidelberg),
  160--172

\bibitem[{{Cantat-Gaudin}(2022)}]{2022Univ....8..111C}
{Cantat-Gaudin}, T. 2022, Universe, 8, 111

\bibitem[{Cantat-Gaudin \& Anders(2020)}]{cantat-gaudin_clusters_mirages_2020}
Cantat-Gaudin, T. \& Anders, F. 2020, Astronomy \& Astrophysics, 633, A99

\bibitem[{{Cantat-Gaudin} {et~al.}(2020){Cantat-Gaudin}, {Anders},
  {Castro-Ginard}, {Jordi}, {Romero-G{\'o}mez}, {Soubiran}, {Casamiquela},
  {Tarricq}, {Moitinho}, {Vallenari}, {Bragaglia}, {Krone-Martins}, \&
  {Kounkel}}]{2020A&A...640A...1C}
{Cantat-Gaudin}, T., {Anders}, F., {Castro-Ginard}, A., {et~al.} 2020, \aap,
  640, A1

\bibitem[{{Cantat-Gaudin} {et~al.}(2016){Cantat-Gaudin}, {Donati}, {Vallenari},
  {Sordo}, {Bragaglia}, \& {Magrini}}]{2016A&A...588A.120C}
{Cantat-Gaudin}, T., {Donati}, P., {Vallenari}, A., {et~al.} 2016, \aap, 588,
  A120

\bibitem[{{Cantat-Gaudin} {et~al.}(2023){Cantat-Gaudin}, {Fouesneau}, {Rix},
  {Brown}, {Castro-Ginard}, {Kostrzewa-Rutkowska}, {Drimmel}, {Hogg}, {Casey},
  {Khanna}, {Oh}, {Price-Whelan}, {Belokurov}, {Saydjari}, \&
  {Green}}]{2023A&A...669A..55C}
{Cantat-Gaudin}, T., {Fouesneau}, M., {Rix}, H.-W., {et~al.} 2023, \aap, 669,
  A55

\bibitem[{{Cantat-Gaudin} {et~al.}(2018){Cantat-Gaudin}, {Jordi}, {Vallenari},
  {Bragaglia}, {Balaguer-N{\'u}{\~n}ez}, {Soubiran}, {Bossini}, {Moitinho},
  {Castro-Ginard}, {Krone-Martins}, {Casamiquela}, {Sordo}, \&
  {Carrera}}]{2018A&A...618A..93C}
{Cantat-Gaudin}, T., {Jordi}, C., {Vallenari}, A., {et~al.} 2018, \aap, 618,
  A93

\bibitem[{{Cantat-Gaudin} {et~al.}(2019){Cantat-Gaudin}, {Krone-Martins},
  {Sedaghat}, {Farahi}, {de Souza}, {Skalidis}, {Malz}, {Mac{\^e}do}, {Moews},
  {Jordi}, {Moitinho}, {Castro-Ginard}, {Ishida}, {Heneka}, {Boucaud}, \&
  {Trindade}}]{2019A&A...624A.126C}
{Cantat-Gaudin}, T., {Krone-Martins}, A., {Sedaghat}, N., {et~al.} 2019, \aap,
  624, A126

\bibitem[{{Carbajo-Hijarrubia} {et~al.}(2024){Carbajo-Hijarrubia},
  {Casamiquela}, {Carrera}, {Balaguer-N{\'u}{\~n}ez}, {Jordi}, {Anders},
  {Gallart}, {Pancino}, {Drazdauskas}, {Stonkut{\.{e}}},
  {Tautvai{\v{s}}ien{\.{e}}}, {Carrasco}, {Masana}, {Cantat-Gaudin}, \&
  {Blanco-Cuaresma}}]{2024A&A...687A.239C}
{Carbajo-Hijarrubia}, J., {Casamiquela}, L., {Carrera}, R., {et~al.} 2024,
  \aap, 687, A239

\bibitem[{{Carraro} {et~al.}(2007){Carraro}, {Geisler}, {Villanova},
  {Frinchaboy}, \& {Majewski}}]{2007A&A...476..217C}
{Carraro}, G., {Geisler}, D., {Villanova}, S., {Frinchaboy}, P.~M., \&
  {Majewski}, S.~R. 2007, \aap, 476, 217

\bibitem[{{Castro-Ginard} {et~al.}(2023){Castro-Ginard}, {Brown},
  {Kostrzewa-Rutkowska}, {Cantat-Gaudin}, {Drimmel}, {Oh}, {Belokurov},
  {Casey}, {Fouesneau}, {Khanna}, {Price-Whelan}, \&
  {Rix}}]{2023A&A...677A..37C}
{Castro-Ginard}, A., {Brown}, A. G.~A., {Kostrzewa-Rutkowska}, Z., {et~al.}
  2023, \aap, 677, A37

\bibitem[{{Castro-Ginard} {et~al.}(2020){Castro-Ginard}, {Jordi}, {Luri},
  {{\'A}lvarez Cid-Fuentes}, {Casamiquela}, {Anders}, {Cantat-Gaudin},
  {Mongui{\'o}}, {Balaguer-N{\'u}{\~n}ez}, {Sol{\`a}}, \&
  {Badia}}]{2020A&A...635A..45C}
{Castro-Ginard}, A., {Jordi}, C., {Luri}, X., {et~al.} 2020, \aap, 635, A45

\bibitem[{{Castro-Ginard} {et~al.}(2018){Castro-Ginard}, {Jordi}, {Luri},
  {Julbe}, {Morvan}, {Balaguer-N{\'u}{\~n}ez}, \&
  {Cantat-Gaudin}}]{2018A&A...618A..59C}
{Castro-Ginard}, A., {Jordi}, C., {Luri}, X., {et~al.} 2018, \aap, 618, A59

\bibitem[{{Cavallo} {et~al.}(2024){Cavallo}, {Spina}, {Carraro}, {Magrini},
  {Poggio}, {Cantat-Gaudin}, {Pasquato}, {Lucatello}, {Ortolani}, \&
  {Schiappacasse-Ulloa}}]{2024AJ....167...12C}
{Cavallo}, L., {Spina}, L., {Carraro}, G., {et~al.} 2024, \aj, 167, 12

\bibitem[{Chen \& Guestrin(2016)}]{xgboost_2016}
Chen, T. \& Guestrin, C. 2016, in Proceedings of the 22nd ACM SIGKDD
  International Conference on Knowledge Discovery and Data Mining, KDD '16 (New
  York, NY, USA: ACM), 785--794

\bibitem[{{Chen} {et~al.}(2019){Chen}, {Wang}, {Deng}, {de Grijs}, {Liu}, \&
  {Tian}}]{2019NatAs...3..320C}
{Chen}, X., {Wang}, S., {Deng}, L., {et~al.} 2019, Nature Astronomy, 3, 320

\bibitem[{{Chen} {et~al.}(2015){Chen}, {Bressan}, {Girardi}, {Marigo}, {Kong},
  \& {Lanza}}]{Chen2015MNRAS.452.1068C}
{Chen}, Y., {Bressan}, A., {Girardi}, L., {et~al.} 2015, \mnras, 452, 1068

\bibitem[{{Chen} {et~al.}(2014){Chen}, {Girardi}, {Bressan}, {Marigo},
  {Barbieri}, \& {Kong}}]{Chen2014MNRAS.444.2525C}
{Chen}, Y., {Girardi}, L., {Bressan}, A., {et~al.} 2014, \mnras, 444, 2525

\bibitem[{{Chen} \& {Zhao}(2020)}]{ChenZhao2020MNRAS.495.2673C}
{Chen}, Y.~Q. \& {Zhao}, G. 2020, \mnras, 495, 2673

\bibitem[{{Dias} {et~al.}(2002){Dias}, {Alessi}, {Moitinho}, \&
  {L{\'e}pine}}]{2002A&A...389..871D}
{Dias}, W.~S., {Alessi}, B.~S., {Moitinho}, A., \& {L{\'e}pine}, J.~R.~D. 2002,
  \aap, 389, 871

\bibitem[{{Donor} {et~al.}(2020){Donor}, {Frinchaboy}, {Cunha}, {O'Connell},
  {Allende Prieto}, {Almeida}, {Anders}, {Beaton}, {Bizyaev}, {Brownstein},
  {Carrera}, {Chiappini}, {Cohen}, {Garc{\'\i}a-Hern{\'a}ndez}, {Geisler},
  {Hasselquist}, {J{\"o}nsson}, {Lane}, {Majewski}, {Minniti}, {Bidin}, {Pan},
  {Roman-Lopes}, {Sobeck}, \& {Zasowski}}]{Donor2020}
{Donor}, J., {Frinchaboy}, P.~M., {Cunha}, K., {et~al.} 2020, \aj, 159, 199

\bibitem[{{Fabricius} {et~al.}(2021){Fabricius}, {Luri}, {Arenou}, {Babusiaux},
  {Helmi}, {Muraveva}, {Reyl{\'e}}, {Spoto}, {Vallenari}, {Antoja}, {Balbinot},
  {Barache}, {Bauchet}, {Bragaglia}, {Busonero}, {Cantat-Gaudin}, {Carrasco},
  {Diakit{\'e}}, {Fabrizio}, {Figueras}, {Garcia-Gutierrez}, {Garofalo},
  {Jordi}, {Kervella}, {Khanna}, {Leclerc}, {Licata}, {Lambert}, {Marrese},
  {Masip}, {Ramos}, {Robichon}, {Robin}, {Romero-G{\'o}mez}, {Rubele}, \&
  {Weiler}}]{Fabricius2021}
{Fabricius}, C., {Luri}, X., {Arenou}, F., {et~al.} 2021, \aap, 649, A5

\bibitem[{{Feast} {et~al.}(2014){Feast}, {Menzies}, {Matsunaga}, \&
  {Whitelock}}]{2014Natur.509..342F}
{Feast}, M.~W., {Menzies}, J.~W., {Matsunaga}, N., \& {Whitelock}, P.~A. 2014,
  \nat, 509, 342

\bibitem[{{Froebrich} {et~al.}(2007){Froebrich}, {Scholz}, \&
  {Raftery}}]{2007MNRAS.374..399F}
{Froebrich}, D., {Scholz}, A., \& {Raftery}, C.~L. 2007, \mnras, 374, 399

\bibitem[{{Gaia Collaboration} {et~al.}(2021{\natexlab{a}}){Gaia
  Collaboration}, {Antoja}, {McMillan}, {Kordopatis}, {Ramos}, {Helmi},
  {Balbinot}, {Cantat-Gaudin}, {Chemin}, {Figueras}, {Jordi}, {Khanna},
  {Romero-G{\'o}mez}, {Seabroke}, {Brown}, {Vallenari}, {Prusti}, {de Bruijne},
  {Babusiaux}, {Biermann}, {Creevey}, {Evans}, {Eyer}, {Hutton}, {Jansen},
  {Klioner}, {Lammers}, {Lindegren}, {Luri}, {Mignard}, {Panem}, {Pourbaix},
  {Randich}, {Sartoretti}, {Soubiran}, {Walton}, {Arenou}, {Bailer-Jones},
  {Bastian}, {Cropper}, {Drimmel}, {Katz}, {Lattanzi}, {van Leeuwen}, {Bakker},
  {Casta{\~n}eda}, {De Angeli}, {Ducourant}, {Fabricius}, {Fouesneau},
  {Fr{\'e}mat}, {Guerra}, {Guerrier}, {Guiraud}, {Jean-Antoine Piccolo},
  {Masana}, {Messineo}, {Mowlavi}, {Nicolas}, {Nienartowicz}, {Pailler},
  {Panuzzo}, {Riclet}, {Roux}, {Sordo}, {Tanga}, {Th{\'e}venin},
  {Gracia-Abril}, {Portell}, {Teyssier}, {Altmann}, {Andrae}, {Bellas-Velidis},
  {Benson}, {Berthier}, {Blomme}, {Brugaletta}, {Burgess}, {Busso}, {Carry},
  {Cellino}, {Cheek}, {Clementini}, {Damerdji}, {Davidson}, {Delchambre},
  {Dell'Oro}, {Fern{\'a}ndez-Hern{\'a}ndez}, {Galluccio}, {Garc{\'\i}a-Lario},
  {Garcia-Reinaldos}, {Gonz{\'a}lez-N{\'u}{\~n}ez}, {Gosset}, {Haigron},
  {Halbwachs}, {Hambly}, {Harrison}, {Hatzidimitriou}, {Heiter},
  {Hern{\'a}ndez}, {Hestroffer}, {Hodgkin}, {Holl}, {Jan{\ss}en}, {Jevardat de
  Fombelle}, {Jordan}, {Krone-Martins}, {Lanzafame}, {L{\"o}ffler}, {Lorca},
  {Manteiga}, {Marchal}, {Marrese}, {Moitinho}, {Mora}, {Muinonen}, {Osborne},
  {Pancino}, {Pauwels}, {Recio-Blanco}, {Richards}, {Riello}, {Rimoldini},
  {Robin}, {Roegiers}, {Rybizki}, {Sarro}, {Siopis}, {Smith}, {Sozzetti},
  {Ulla}, {Utrilla}, {van Leeuwen}, {van Reeven}, {Abbas}, {Abreu Aramburu},
  {Accart}, {Aerts}, {Aguado}, {Ajaj}, {Altavilla}, {{\'A}lvarez}, {{\'A}lvarez
  Cid-Fuentes}, {Alves}, {Anderson}, {Varela}, {Audard}, {Baines}, {Baker},
  {Balaguer-N{\'u}{\~n}ez}, {Balog}, {Barache}, {Barbato}, {Barros}, {Barstow},
  {Bartolom{\'e}}, {Bassilana}, {Bauchet}, {Baudesson-Stella}, {Becciani},
  {Bellazzini}, {Bernet}, {Bertone}, {Bianchi}, {Blanco-Cuaresma}, {Boch},
  {Bombrun}, {Bossini}, {Bouquillon}, {Bragaglia}, {Bramante}, {Breedt},
  {Bressan}, {Brouillet}, {Bucciarelli}, {Burlacu}, {Busonero}, {Butkevich},
  {Buzzi}, {Caffau}, {Cancelliere}, {C{\'a}novas}, {Carballo}, {Carlucci},
  {Carnerero}, {Carrasco}, {Casamiquela}, {Castellani}, {Castro-Ginard},
  {Castro Sampol}, {Chaoul}, {Charlot}, {Chiavassa}, {Cioni}, {Comoretto},
  {Cooper}, {Cornez}, {Cowell}, {Crifo}, {Crosta}, {Crowley}, {Dafonte},
  {Dapergolas}, {David}, {David}, {de Laverny}, {De Luise}, {De March}, {De
  Ridder}, {de Souza}, {de Teodoro}, {de Torres}, {del Peloso}, {del Pozo},
  {Delgado}, {Delgado}, {Delisle}, {Di Matteo}, {Diakite}, {Diener},
  {Distefano}, {Dolding}, {Eappachen}, {Enke}, {Esquej}, {Fabre}, {Fabrizio},
  {Faigler}, {Fedorets}, {Fernique}, {Fienga}, {Fouron}, {Fragkoudi}, {Fraile},
  {Franke}, {Gai}, {Garabato}, {Garcia-Gutierrez}, {Garc{\'\i}a-Torres},
  {Garofalo}, {Gavras}, {Gerlach}, {Geyer}, {Giacobbe}, {Gilmore}, {Girona},
  {Giuffrida}, {Gomez}, {Gonzalez-Santamaria}, {Gonz{\'a}lez-Vidal}, {Granvik},
  {Guti{\'e}rrez-S{\'a}nchez}, {Guy}, {Hauser}, {Haywood}, {Hidalgo}, {Hilger},
  {H{\l}adczuk}, {Hobbs}, {Holland}, {Huckle}, {Jasniewicz}, {Jonker},
  {Juaristi Campillo}, {Julbe}, {Karbevska}, {Kervella}, {Kochoska},
  {Kontizas}, {Korn}, {Kostrzewa-Rutkowska}, {Kruszy{\'n}ska}, {Lambert},
  {Lanza}, {Lasne}, {Le Campion}, {Le Fustec}, {Lebreton}, {Lebzelter},
  {Leccia}, {Leclerc}, {Lecoeur-Taibi}, {Liao}, {Licata}, {Lindstr{\o}m},
  {Lister}, {Livanou}, {Lobel}, {Madrero Pardo}, {Managau}, {Mann}, {Marchant},
  {Marconi}, {Marcos Santos}, {Marinoni}, {Marocco}, {Marshall}, {Martin Polo},
  {Mart{\'\i}n-Fleitas}, {Masip}, {Massari}, {Mastrobuono-Battisti}, {Mazeh},
  {Messina}, {Michalik}, {Millar}, {Mints}, {Molina}, {Molinaro}, {Moln{\'a}r},
  {Montegriffo}, {Mor}, {Morbidelli}, {Morel}, {Morris}, {Mulone}, {Munoz},
  {Muraveva}, {Murphy}, {Musella}, {Noval}, {Ord{\'e}novic}, {Orr{\`u}},
  {Osinde}, {Pagani}, {Pagano}, {Palaversa}, {Palicio}, {Panahi}, {Pawlak},
  {Pe{\~n}alosa Esteller}, {Penttil{\"a}}, {Piersimoni}, {Pineau}, {Plachy},
  {Plum}, {Poggio}, {Poretti}, {Poujoulet}, {Pr{\v{s}}a}, {Pulone}, {Racero},
  {Ragaini}, {Rainer}, {Raiteri}, {Rambaux}, {Ramos-Lerate}, {Re Fiorentin},
  {Regibo}, {Reyl{\'e}}, {Ripepi}, {Riva}, {Rixon}, {Robichon}, {Robin},
  {Roelens}, {Rohrbasser}, {Rowell}, {Royer}, {Rybicki}, {Sadowski},
  {Sagrist{\`a} Sell{\'e}s}, {Sahlmann}, {Salgado}, {Salguero}, {Samaras},
  {Sanchez Gimenez}, {Sanna}, {Santove{\~n}a}, {Sarasso}, {Schultheis},
  {Sciacca}, {Segol}, {Segovia}, {S{\'e}gransan}, {Semeux}, {Siddiqui},
  {Siebert}, {Siltala}, {Slezak}, {Smart}, {Solano}, {Solitro}, {Souami},
  {Souchay}, {Spagna}, {Spoto}, {Steele}, {Steidelm{\"u}ller}, {Stephenson},
  {S{\"u}veges}, {Szabados}, {Szegedi-Elek}, {Taris}, {Tauran}, {Taylor},
  {Teixeira}, {Thuillot}, {Tonello}, {Torra}, {Torra}, {Turon}, {Unger},
  {Vaillant}, {van Dillen}, {Vanel}, {Vecchiato}, {Viala}, {Vicente},
  {Voutsinas}, {Weiler}, {Wevers}, {Wyrzykowski}, {Yoldas}, {Yvard}, {Zhao},
  {Zorec}, {Zucker}, {Zurbach}, \& {Zwitter}}]{GaiaEDR3Anticentre}
{Gaia Collaboration}, {Antoja}, T., {McMillan}, P.~J., {et~al.}
  2021{\natexlab{a}}, \aap, 649, A8

\bibitem[{{Gaia Collaboration} {et~al.}(2018){Gaia Collaboration}, {Brown},
  {Vallenari}, {Prusti}, {de Bruijne}, {Babusiaux}, {Bailer-Jones}, {Biermann},
  {Evans}, {Eyer}, {Jansen}, {Jordi}, {Klioner}, {Lammers}, {Lindegren},
  {Luri}, {Mignard}, {Panem}, {Pourbaix}, {Randich}, {Sartoretti}, {Siddiqui},
  {Soubiran}, {van Leeuwen}, {Walton}, {Arenou}, {Bastian}, {Cropper},
  {Drimmel}, {Katz}, {Lattanzi}, {Bakker}, {Cacciari}, {Casta{\~n}eda},
  {Chaoul}, {Cheek}, {De Angeli}, {Fabricius}, {Guerra}, {Holl}, {Masana},
  {Messineo}, {Mowlavi}, {Nienartowicz}, {Panuzzo}, {Portell}, {Riello},
  {Seabroke}, {Tanga}, {Th{\'e}venin}, {Gracia-Abril}, {Comoretto},
  {Garcia-Reinaldos}, {Teyssier}, {Altmann}, {Andrae}, {Audard},
  {Bellas-Velidis}, {Benson}, {Berthier}, {Blomme}, {Burgess}, {Busso},
  {Carry}, {Cellino}, {Clementini}, {Clotet}, {Creevey}, {Davidson}, {De
  Ridder}, {Delchambre}, {Dell'Oro}, {Ducourant},
  {Fern{\'a}ndez-Hern{\'a}ndez}, {Fouesneau}, {Fr{\'e}mat}, {Galluccio},
  {Garc{\'\i}a-Torres}, {Gonz{\'a}lez-N{\'u}{\~n}ez}, {Gonz{\'a}lez-Vidal},
  {Gosset}, {Guy}, {Halbwachs}, {Hambly}, {Harrison}, {Hern{\'a}ndez},
  {Hestroffer}, {Hodgkin}, {Hutton}, {Jasniewicz}, {Jean-Antoine-Piccolo},
  {Jordan}, {Korn}, {Krone-Martins}, {Lanzafame}, {Lebzelter}, {L{\"o}ffler},
  {Manteiga}, {Marrese}, {Mart{\'\i}n-Fleitas}, {Moitinho}, {Mora}, {Muinonen},
  {Osinde}, {Pancino}, {Pauwels}, {Petit}, {Recio-Blanco}, {Richards},
  {Rimoldini}, {Robin}, {Sarro}, {Siopis}, {Smith}, {Sozzetti}, {S{\"u}veges},
  {Torra}, {van Reeven}, {Abbas}, {Abreu Aramburu}, {Accart}, {Aerts},
  {Altavilla}, {{\'A}lvarez}, {Alvarez}, {Alves}, {Anderson}, {Andrei},
  {Anglada Varela}, {Antiche}, {Antoja}, {Arcay}, {Astraatmadja}, {Bach},
  {Baker}, {Balaguer-N{\'u}{\~n}ez}, {Balm}, {Barache}, {Barata}, {Barbato},
  {Barblan}, {Barklem}, {Barrado}, {Barros}, {Barstow}, {Bartholom{\'e}
  Mu{\~n}oz}, {Bassilana}, {Becciani}, {Bellazzini}, {Berihuete}, {Bertone},
  {Bianchi}, {Bienaym{\'e}}, {Blanco-Cuaresma}, {Boch}, {Boeche}, {Bombrun},
  {Borrachero}, {Bossini}, {Bouquillon}, {Bourda}, {Bragaglia}, {Bramante},
  {Breddels}, {Bressan}, {Brouillet}, {Br{\"u}semeister}, {Brugaletta},
  {Bucciarelli}, {Burlacu}, {Busonero}, {Butkevich}, {Buzzi}, {Caffau},
  {Cancelliere}, {Cannizzaro}, {Cantat-Gaudin}, {Carballo}, {Carlucci},
  {Carrasco}, {Casamiquela}, {Castellani}, {Castro-Ginard}, {Charlot},
  {Chemin}, {Chiavassa}, {Cocozza}, {Costigan}, {Cowell}, {Crifo}, {Crosta},
  {Crowley}, {Cuypers}, {Dafonte}, {Damerdji}, {Dapergolas}, {David}, {David},
  {de Laverny}, {De Luise}, {De March}, {de Martino}, {de Souza}, {de Torres},
  {Debosscher}, {del Pozo}, {Delbo}, {Delgado}, {Delgado}, {Di Matteo},
  {Diakite}, {Diener}, {Distefano}, {Dolding}, {Drazinos}, {Dur{\'a}n},
  {Edvardsson}, {Enke}, {Eriksson}, {Esquej}, {Eynard Bontemps}, {Fabre},
  {Fabrizio}, {Faigler}, {Falc{\~a}o}, {Farr{\`a}s Casas}, {Federici},
  {Fedorets}, {Fernique}, {Figueras}, {Filippi}, {Findeisen}, {Fonti},
  {Fraile}, {Fraser}, {Fr{\'e}zouls}, {Gai}, {Galleti}, {Garabato},
  {Garc{\'\i}a-Sedano}, {Garofalo}, {Garralda}, {Gavel}, {Gavras}, {Gerssen},
  {Geyer}, {Giacobbe}, {Gilmore}, {Girona}, {Giuffrida}, {Glass}, {Gomes},
  {Granvik}, {Gueguen}, {Guerrier}, {Guiraud}, {Guti{\'e}rrez-S{\'a}nchez},
  {Haigron}, {Hatzidimitriou}, {Hauser}, {Haywood}, {Heiter}, {Helmi}, {Heu},
  {Hilger}, {Hobbs}, {Hofmann}, {Holland}, {Huckle}, {Hypki}, {Icardi},
  {Jan{\ss}en}, {Jevardat de Fombelle}, {Jonker}, {Juh{\'a}sz}, {Julbe},
  {Karampelas}, {Kewley}, {Klar}, {Kochoska}, {Kohley}, {Kolenberg},
  {Kontizas}, {Kontizas}, {Koposov}, {Kordopatis}, {Kostrzewa-Rutkowska},
  {Koubsky}, {Lambert}, {Lanza}, {Lasne}, {Lavigne}, {Le Fustec}, {Le
  Poncin-Lafitte}, {Lebreton}, {Leccia}, {Leclerc}, {Lecoeur-Taibi},
  {Lenhardt}, {Leroux}, {Liao}, {Licata}, {Lindstr{\o}m}, {Lister}, {Livanou},
  {Lobel}, {L{\'o}pez}, {Managau}, {Mann}, {Mantelet}, {Marchal}, {Marchant},
  {Marconi}, {Marinoni}, {Marschalk{\'o}}, {Marshall}, {Martino}, {Marton},
  {Mary}, {Massari}, {Matijevi{\v{c}}}, {Mazeh}, {McMillan}, {Messina},
  {Michalik}, {Millar}, {Molina}, {Molinaro}, {Moln{\'a}r}, {Montegriffo},
  {Mor}, {Morbidelli}, {Morel}, {Morris}, {Mulone}, {Muraveva}, {Musella},
  {Nelemans}, {Nicastro}, {Noval}, {O'Mullane}, {Ord{\'e}novic},
  {Ord{\'o}{\~n}ez-Blanco}, {Osborne}, {Pagani}, {Pagano}, {Pailler},
  {Palacin}, {Palaversa}, {Panahi}, {Pawlak}, {Piersimoni}, {Pineau}, {Plachy},
  {Plum}, {Poggio}, {Poujoulet}, {Pr{\v{s}}a}, {Pulone}, {Racero}, {Ragaini},
  {Rambaux}, {Ramos-Lerate}, {Regibo}, {Reyl{\'e}}, {Riclet}, {Ripepi}, {Riva},
  {Rivard}, {Rixon}, {Roegiers}, {Roelens}, {Romero-G{\'o}mez}, {Rowell},
  {Royer}, {Ruiz-Dern}, {Sadowski}, {Sagrist{\`a} Sell{\'e}s}, {Sahlmann},
  {Salgado}, {Salguero}, {Sanna}, {Santana-Ros}, {Sarasso}, {Savietto},
  {Schultheis}, {Sciacca}, {Segol}, {Segovia}, {S{\'e}gransan}, {Shih},
  {Siltala}, {Silva}, {Smart}, {Smith}, {Solano}, {Solitro}, {Sordo}, {Soria
  Nieto}, {Souchay}, {Spagna}, {Spoto}, {Stampa}, {Steele},
  {Steidelm{\"u}ller}, {Stephenson}, {Stoev}, {Suess}, {Surdej}, {Szabados},
  {Szegedi-Elek}, {Tapiador}, {Taris}, {Tauran}, {Taylor}, {Teixeira},
  {Terrett}, {Teyssandier}, {Thuillot}, {Titarenko}, {Torra Clotet}, {Turon},
  {Ulla}, {Utrilla}, {Uzzi}, {Vaillant}, {Valentini}, {Valette}, {van Elteren},
  {Van Hemelryck}, {van Leeuwen}, {Vaschetto}, {Vecchiato}, {Veljanoski},
  {Viala}, {Vicente}, {Vogt}, {von Essen}, {Voss}, {Votruba}, {Voutsinas},
  {Walmsley}, {Weiler}, {Wertz}, {Wevers}, {Wyrzykowski}, {Yoldas},
  {{\v{Z}}erjal}, {Ziaeepour}, {Zorec}, {Zschocke}, {Zucker}, {Zurbach}, \&
  {Zwitter}}]{GaiaDR2}
{Gaia Collaboration}, {Brown}, A.~G.~A., {Vallenari}, A., {et~al.} 2018, \aap,
  616, A1

\bibitem[{{Gaia Collaboration} {et~al.}(2021{\natexlab{b}}){Gaia
  Collaboration}, {Brown}, {Vallenari}, {Prusti}, {de Bruijne}, {Babusiaux},
  {Biermann}, {Creevey}, {Evans}, {Eyer}, {Hutton}, {Jansen}, {Jordi},
  {Klioner}, {Lammers}, {Lindegren}, {Luri}, {Mignard}, {Panem}, {Pourbaix},
  {Randich}, {Sartoretti}, {Soubiran}, {Walton}, {Arenou}, {Bailer-Jones},
  {Bastian}, {Cropper}, {Drimmel}, {Katz}, {Lattanzi}, {van Leeuwen}, {Bakker},
  {Cacciari}, {Casta{\~n}eda}, {De Angeli}, {Ducourant}, {Fabricius},
  {Fouesneau}, {Fr{\'e}mat}, {Guerra}, {Guerrier}, {Guiraud}, {Jean-Antoine
  Piccolo}, {Masana}, {Messineo}, {Mowlavi}, {Nicolas}, {Nienartowicz},
  {Pailler}, {Panuzzo}, {Riclet}, {Roux}, {Seabroke}, {Sordo}, {Tanga},
  {Th{\'e}venin}, {Gracia-Abril}, {Portell}, {Teyssier}, {Altmann}, {Andrae},
  {Bellas-Velidis}, {Benson}, {Berthier}, {Blomme}, {Brugaletta}, {Burgess},
  {Busso}, {Carry}, {Cellino}, {Cheek}, {Clementini}, {Damerdji}, {Davidson},
  {Delchambre}, {Dell'Oro}, {Fern{\'a}ndez-Hern{\'a}ndez}, {Galluccio},
  {Garc{\'\i}a-Lario}, {Garcia-Reinaldos}, {Gonz{\'a}lez-N{\'u}{\~n}ez},
  {Gosset}, {Haigron}, {Halbwachs}, {Hambly}, {Harrison}, {Hatzidimitriou},
  {Heiter}, {Hern{\'a}ndez}, {Hestroffer}, {Hodgkin}, {Holl}, {Jan{\ss}en},
  {Jevardat de Fombelle}, {Jordan}, {Krone-Martins}, {Lanzafame},
  {L{\"o}ffler}, {Lorca}, {Manteiga}, {Marchal}, {Marrese}, {Moitinho}, {Mora},
  {Muinonen}, {Osborne}, {Pancino}, {Pauwels}, {Petit}, {Recio-Blanco},
  {Richards}, {Riello}, {Rimoldini}, {Robin}, {Roegiers}, {Rybizki}, {Sarro},
  {Siopis}, {Smith}, {Sozzetti}, {Ulla}, {Utrilla}, {van Leeuwen}, {van
  Reeven}, {Abbas}, {Abreu Aramburu}, {Accart}, {Aerts}, {Aguado}, {Ajaj},
  {Altavilla}, {{\'A}lvarez}, {{\'A}lvarez Cid-Fuentes}, {Alves}, {Anderson},
  {Anglada Varela}, {Antoja}, {Audard}, {Baines}, {Baker},
  {Balaguer-N{\'u}{\~n}ez}, {Balbinot}, {Balog}, {Barache}, {Barbato},
  {Barros}, {Barstow}, {Bartolom{\'e}}, {Bassilana}, {Bauchet},
  {Baudesson-Stella}, {Becciani}, {Bellazzini}, {Bernet}, {Bertone}, {Bianchi},
  {Blanco-Cuaresma}, {Boch}, {Bombrun}, {Bossini}, {Bouquillon}, {Bragaglia},
  {Bramante}, {Breedt}, {Bressan}, {Brouillet}, {Bucciarelli}, {Burlacu},
  {Busonero}, {Butkevich}, {Buzzi}, {Caffau}, {Cancelliere}, {C{\'a}novas},
  {Cantat-Gaudin}, {Carballo}, {Carlucci}, {Carnerero}, {Carrasco},
  {Casamiquela}, {Castellani}, {Castro-Ginard}, {Castro Sampol}, {Chaoul},
  {Charlot}, {Chemin}, {Chiavassa}, {Cioni}, {Comoretto}, {Cooper}, {Cornez},
  {Cowell}, {Crifo}, {Crosta}, {Crowley}, {Dafonte}, {Dapergolas}, {David},
  {David}, {de Laverny}, {De Luise}, {De March}, {De Ridder}, {de Souza}, {de
  Teodoro}, {de Torres}, {del Peloso}, {del Pozo}, {Delbo}, {Delgado},
  {Delgado}, {Delisle}, {Di Matteo}, {Diakite}, {Diener}, {Distefano},
  {Dolding}, {Eappachen}, {Edvardsson}, {Enke}, {Esquej}, {Fabre}, {Fabrizio},
  {Faigler}, {Fedorets}, {Fernique}, {Fienga}, {Figueras}, {Fouron},
  {Fragkoudi}, {Fraile}, {Franke}, {Gai}, {Garabato}, {Garcia-Gutierrez},
  {Garc{\'\i}a-Torres}, {Garofalo}, {Gavras}, {Gerlach}, {Geyer}, {Giacobbe},
  {Gilmore}, {Girona}, {Giuffrida}, {Gomel}, {Gomez}, {Gonzalez-Santamaria},
  {Gonz{\'a}lez-Vidal}, {Granvik}, {Guti{\'e}rrez-S{\'a}nchez}, {Guy},
  {Hauser}, {Haywood}, {Helmi}, {Hidalgo}, {Hilger}, {H{\l}adczuk}, {Hobbs},
  {Holland}, {Huckle}, {Jasniewicz}, {Jonker}, {Juaristi Campillo}, {Julbe},
  {Karbevska}, {Kervella}, {Khanna}, {Kochoska}, {Kontizas}, {Kordopatis},
  {Korn}, {Kostrzewa-Rutkowska}, {Kruszy{\'n}ska}, {Lambert}, {Lanza}, {Lasne},
  {Le Campion}, {Le Fustec}, {Lebreton}, {Lebzelter}, {Leccia}, {Leclerc},
  {Lecoeur-Taibi}, {Liao}, {Licata}, {Lindstr{\o}m}, {Lister}, {Livanou},
  {Lobel}, {Madrero Pardo}, {Managau}, {Mann}, {Marchant}, {Marconi}, {Marcos
  Santos}, {Marinoni}, {Marocco}, {Marshall}, {Martin Polo},
  {Mart{\'\i}n-Fleitas}, {Masip}, {Massari}, {Mastrobuono-Battisti}, {Mazeh},
  {McMillan}, {Messina}, {Michalik}, {Millar}, {Mints}, {Molina}, {Molinaro},
  {Moln{\'a}r}, {Montegriffo}, {Mor}, {Morbidelli}, {Morel}, {Morris},
  {Mulone}, {Munoz}, {Muraveva}, {Murphy}, {Musella}, {Noval}, {Ord{\'e}novic},
  {Orr{\`u}}, {Osinde}, {Pagani}, {Pagano}, {Palaversa}, {Palicio}, {Panahi},
  {Pawlak}, {Pe{\~n}alosa Esteller}, {Penttil{\"a}}, {Piersimoni}, {Pineau},
  {Plachy}, {Plum}, {Poggio}, {Poretti}, {Poujoulet}, {Pr{\v{s}}a}, {Pulone},
  {Racero}, {Ragaini}, {Rainer}, {Raiteri}, {Rambaux}, {Ramos}, {Ramos-Lerate},
  {Re Fiorentin}, {Regibo}, {Reyl{\'e}}, {Ripepi}, {Riva}, {Rixon}, {Robichon},
  {Robin}, {Roelens}, {Rohrbasser}, {Romero-G{\'o}mez}, {Rowell}, {Royer},
  {Rybicki}, {Sadowski}, {Sagrist{\`a} Sell{\'e}s}, {Sahlmann}, {Salgado},
  {Salguero}, {Samaras}, {Sanchez Gimenez}, {Sanna}, {Santove{\~n}a},
  {Sarasso}, {Schultheis}, {Sciacca}, {Segol}, {Segovia}, {S{\'e}gransan},
  {Semeux}, {Shahaf}, {Siddiqui}, {Siebert}, {Siltala}, {Slezak}, {Smart},
  {Solano}, {Solitro}, {Souami}, {Souchay}, {Spagna}, {Spoto}, {Steele},
  {Steidelm{\"u}ller}, {Stephenson}, {S{\"u}veges}, {Szabados}, {Szegedi-Elek},
  {Taris}, {Tauran}, {Taylor}, {Teixeira}, {Thuillot}, {Tonello}, {Torra},
  {Torra}, {Turon}, {Unger}, {Vaillant}, {van Dillen}, {Vanel}, {Vecchiato},
  {Viala}, {Vicente}, {Voutsinas}, {Weiler}, {Wevers}, {Wyrzykowski}, {Yoldas},
  {Yvard}, {Zhao}, {Zorec}, {Zucker}, {Zurbach}, \& {Zwitter}}]{GaiaDR3}
{Gaia Collaboration}, {Brown}, A.~G.~A., {Vallenari}, A., {et~al.}
  2021{\natexlab{b}}, \aap, 649, A1

\bibitem[{{Gaia Collaboration} {et~al.}(2016){Gaia Collaboration}, {Prusti},
  {de Bruijne}, {Brown}, {Vallenari}, {Babusiaux}, {Bailer-Jones}, {Bastian},
  {Biermann}, {Evans}, {Eyer}, {Jansen}, {Jordi}, {Klioner}, {Lammers},
  {Lindegren}, {Luri}, {Mignard}, {Milligan}, {Panem}, {Poinsignon},
  {Pourbaix}, {Randich}, {Sarri}, {Sartoretti}, {Siddiqui}, {Soubiran},
  {Valette}, {van Leeuwen}, {Walton}, {Aerts}, {Arenou}, {Cropper}, {Drimmel},
  {H{\o}g}, {Katz}, {Lattanzi}, {O'Mullane}, {Grebel}, {Holland}, {Huc},
  {Passot}, {Bramante}, {Cacciari}, {Casta{\~n}eda}, {Chaoul}, {Cheek}, {De
  Angeli}, {Fabricius}, {Guerra}, {Hern{\'a}ndez}, {Jean-Antoine-Piccolo},
  {Masana}, {Messineo}, {Mowlavi}, {Nienartowicz}, {Ord{\'o}{\~n}ez-Blanco},
  {Panuzzo}, {Portell}, {Richards}, {Riello}, {Seabroke}, {Tanga},
  {Th{\'e}venin}, {Torra}, {Els}, {Gracia-Abril}, {Comoretto},
  {Garcia-Reinaldos}, {Lock}, {Mercier}, {Altmann}, {Andrae}, {Astraatmadja},
  {Bellas-Velidis}, {Benson}, {Berthier}, {Blomme}, {Busso}, {Carry},
  {Cellino}, {Clementini}, {Cowell}, {Creevey}, {Cuypers}, {Davidson}, {De
  Ridder}, {de Torres}, {Delchambre}, {Dell'Oro}, {Ducourant}, {Fr{\'e}mat},
  {Garc{\'\i}a-Torres}, {Gosset}, {Halbwachs}, {Hambly}, {Harrison}, {Hauser},
  {Hestroffer}, {Hodgkin}, {Huckle}, {Hutton}, {Jasniewicz}, {Jordan},
  {Kontizas}, {Korn}, {Lanzafame}, {Manteiga}, {Moitinho}, {Muinonen},
  {Osinde}, {Pancino}, {Pauwels}, {Petit}, {Recio-Blanco}, {Robin}, {Sarro},
  {Siopis}, {Smith}, {Smith}, {Sozzetti}, {Thuillot}, {van Reeven}, {Viala},
  {Abbas}, {Abreu Aramburu}, {Accart}, {Aguado}, {Allan}, {Allasia},
  {Altavilla}, {{\'A}lvarez}, {Alves}, {Anderson}, {Andrei}, {Anglada Varela},
  {Antiche}, {Antoja}, {Ant{\'o}n}, {Arcay}, {Atzei}, {Ayache}, {Bach},
  {Baker}, {Balaguer-N{\'u}{\~n}ez}, {Barache}, {Barata}, {Barbier}, {Barblan},
  {Baroni}, {Barrado y Navascu{\'e}s}, {Barros}, {Barstow}, {Becciani},
  {Bellazzini}, {Bellei}, {Bello Garc{\'\i}a}, {Belokurov}, {Bendjoya},
  {Berihuete}, {Bianchi}, {Bienaym{\'e}}, {Billebaud}, {Blagorodnova},
  {Blanco-Cuaresma}, {Boch}, {Bombrun}, {Borrachero}, {Bouquillon}, {Bourda},
  {Bouy}, {Bragaglia}, {Breddels}, {Brouillet}, {Br{\"u}semeister},
  {Bucciarelli}, {Budnik}, {Burgess}, {Burgon}, {Burlacu}, {Busonero}, {Buzzi},
  {Caffau}, {Cambras}, {Campbell}, {Cancelliere}, {Cantat-Gaudin}, {Carlucci},
  {Carrasco}, {Castellani}, {Charlot}, {Charnas}, {Charvet}, {Chassat},
  {Chiavassa}, {Clotet}, {Cocozza}, {Collins}, {Collins}, {Costigan}, {Crifo},
  {Cross}, {Crosta}, {Crowley}, {Dafonte}, {Damerdji}, {Dapergolas}, {David},
  {David}, {De Cat}, {de Felice}, {de Laverny}, {De Luise}, {De March}, {de
  Martino}, {de Souza}, {Debosscher}, {del Pozo}, {Delbo}, {Delgado},
  {Delgado}, {di Marco}, {Di Matteo}, {Diakite}, {Distefano}, {Dolding}, {Dos
  Anjos}, {Drazinos}, {Dur{\'a}n}, {Dzigan}, {Ecale}, {Edvardsson}, {Enke},
  {Erdmann}, {Escolar}, {Espina}, {Evans}, {Eynard Bontemps}, {Fabre},
  {Fabrizio}, {Faigler}, {Falc{\~a}o}, {Farr{\`a}s Casas}, {Faye}, {Federici},
  {Fedorets}, {Fern{\'a}ndez-Hern{\'a}ndez}, {Fernique}, {Fienga}, {Figueras},
  {Filippi}, {Findeisen}, {Fonti}, {Fouesneau}, {Fraile}, {Fraser}, {Fuchs},
  {Furnell}, {Gai}, {Galleti}, {Galluccio}, {Garabato}, {Garc{\'\i}a-Sedano},
  {Gar{\'e}}, {Garofalo}, {Garralda}, {Gavras}, {Gerssen}, {Geyer}, {Gilmore},
  {Girona}, {Giuffrida}, {Gomes}, {Gonz{\'a}lez-Marcos},
  {Gonz{\'a}lez-N{\'u}{\~n}ez}, {Gonz{\'a}lez-Vidal}, {Granvik}, {Guerrier},
  {Guillout}, {Guiraud}, {G{\'u}rpide}, {Guti{\'e}rrez-S{\'a}nchez}, {Guy},
  {Haigron}, {Hatzidimitriou}, {Haywood}, {Heiter}, {Helmi}, {Hobbs},
  {Hofmann}, {Holl}, {Holland}, {Hunt}, {Hypki}, {Icardi}, {Irwin}, {Jevardat
  de Fombelle}, {Jofr{\'e}}, {Jonker}, {Jorissen}, {Julbe}, {Karampelas},
  {Kochoska}, {Kohley}, {Kolenberg}, {Kontizas}, {Koposov}, {Kordopatis},
  {Koubsky}, {Kowalczyk}, {Krone-Martins}, {Kudryashova}, {Kull}, {Bachchan},
  {Lacoste-Seris}, {Lanza}, {Lavigne}, {Le Poncin-Lafitte}, {Lebreton},
  {Lebzelter}, {Leccia}, {Leclerc}, {Lecoeur-Taibi}, {Lemaitre}, {Lenhardt},
  {Leroux}, {Liao}, {Licata}, {Lindstr{\o}m}, {Lister}, {Livanou}, {Lobel},
  {L{\"o}ffler}, {L{\'o}pez}, {Lopez-Lozano}, {Lorenz}, {Loureiro},
  {MacDonald}, {Magalh{\~a}es Fernandes}, {Managau}, {Mann}, {Mantelet},
  {Marchal}, {Marchant}, {Marconi}, {Marie}, {Marinoni}, {Marrese},
  {Marschalk{\'o}}, {Marshall}, {Mart{\'\i}n-Fleitas}, {Martino}, {Mary},
  {Matijevi{\v{c}}}, {Mazeh}, {McMillan}, {Messina}, {Mestre}, {Michalik},
  {Millar}, {Miranda}, {Molina}, {Molinaro}, {Molinaro}, {Moln{\'a}r},
  {Moniez}, {Montegriffo}, {Monteiro}, {Mor}, {Mora}, {Morbidelli}, {Morel},
  {Morgenthaler}, {Morley}, {Morris}, {Mulone}, {Muraveva}, {Musella},
  {Narbonne}, {Nelemans}, {Nicastro}, {Noval}, {Ord{\'e}novic},
  {Ordieres-Mer{\'e}}, {Osborne}, {Pagani}, {Pagano}, {Pailler}, {Palacin},
  {Palaversa}, {Parsons}, {Paulsen}, {Pecoraro}, {Pedrosa}, {Pentik{\"a}inen},
  {Pereira}, {Pichon}, {Piersimoni}, {Pineau}, {Plachy}, {Plum}, {Poujoulet},
  {Pr{\v{s}}a}, {Pulone}, {Ragaini}, {Rago}, {Rambaux}, {Ramos-Lerate},
  {Ranalli}, {Rauw}, {Read}, {Regibo}, {Renk}, {Reyl{\'e}}, {Ribeiro},
  {Rimoldini}, {Ripepi}, {Riva}, {Rixon}, {Roelens}, {Romero-G{\'o}mez},
  {Rowell}, {Royer}, {Rudolph}, {Ruiz-Dern}, {Sadowski}, {Sagrist{\`a}
  Sell{\'e}s}, {Sahlmann}, {Salgado}, {Salguero}, {Sarasso}, {Savietto},
  {Schnorhk}, {Schultheis}, {Sciacca}, {Segol}, {Segovia}, {Segransan},
  {Serpell}, {Shih}, {Smareglia}, {Smart}, {Smith}, {Solano}, {Solitro},
  {Sordo}, {Soria Nieto}, {Souchay}, {Spagna}, {Spoto}, {Stampa}, {Steele},
  {Steidelm{\"u}ller}, {Stephenson}, {Stoev}, {Suess}, {S{\"u}veges}, {Surdej},
  {Szabados}, {Szegedi-Elek}, {Tapiador}, {Taris}, {Tauran}, {Taylor},
  {Teixeira}, {Terrett}, {Tingley}, {Trager}, {Turon}, {Ulla}, {Utrilla},
  {Valentini}, {van Elteren}, {Van Hemelryck}, {van Leeuwen}, {Varadi},
  {Vecchiato}, {Veljanoski}, {Via}, {Vicente}, {Vogt}, {Voss}, {Votruba},
  {Voutsinas}, {Walmsley}, {Weiler}, {Weingrill}, {Werner}, {Wevers},
  {Whitehead}, {Wyrzykowski}, {Yoldas}, {{\v{Z}}erjal}, {Zucker}, {Zurbach},
  {Zwitter}, {Alecu}, {Allen}, {Allende Prieto}, {Amorim},
  {Anglada-Escud{\'e}}, {Arsenijevic}, {Azaz}, {Balm}, {Beck}, {Bernstein},
  {Bigot}, {Bijaoui}, {Blasco}, {Bonfigli}, {Bono}, {Boudreault}, {Bressan},
  {Brown}, {Brunet}, {Bunclark}, {Buonanno}, {Butkevich}, {Carret}, {Carrion},
  {Chemin}, {Ch{\'e}reau}, {Corcione}, {Darmigny}, {de Boer}, {de Teodoro}, {de
  Zeeuw}, {Delle Luche}, {Domingues}, {Dubath}, {Fodor}, {Fr{\'e}zouls},
  {Fries}, {Fustes}, {Fyfe}, {Gallardo}, {Gallegos}, {Gardiol}, {Gebran},
  {Gomboc}, {G{\'o}mez}, {Grux}, {Gueguen}, {Heyrovsky}, {Hoar}, {Iannicola},
  {Isasi Parache}, {Janotto}, {Joliet}, {Jonckheere}, {Keil}, {Kim},
  {Klagyivik}, {Klar}, {Knude}, {Kochukhov}, {Kolka}, {Kos}, {Kutka}, {Lainey},
  {LeBouquin}, {Liu}, {Loreggia}, {Makarov}, {Marseille}, {Martayan},
  {Martinez-Rubi}, {Massart}, {Meynadier}, {Mignot}, {Munari}, {Nguyen},
  {Nordlander}, {Ocvirk}, {O'Flaherty}, {Olias Sanz}, {Ortiz}, {Osorio},
  {Oszkiewicz}, {Ouzounis}, {Palmer}, {Park}, {Pasquato}, {Peltzer}, {Peralta},
  {P{\'e}turaud}, {Pieniluoma}, {Pigozzi}, {Poels}, {Prat}, {Prod'homme},
  {Raison}, {Rebordao}, {Risquez}, {Rocca-Volmerange}, {Rosen}, {Ruiz-Fuertes},
  {Russo}, {Sembay}, {Serraller Vizcaino}, {Short}, {Siebert}, {Silva},
  {Sinachopoulos}, {Slezak}, {Soffel}, {Sosnowska}, {Strai{\v{z}}ys}, {ter
  Linden}, {Terrell}, {Theil}, {Tiede}, {Troisi}, {Tsalmantza}, {Tur},
  {Vaccari}, {Vachier}, {Valles}, {Van Hamme}, {Veltz}, {Virtanen}, {Wallut},
  {Wichmann}, {Wilkinson}, {Ziaeepour}, \& {Zschocke}}]{2016A&A...595A...1G}
{Gaia Collaboration}, {Prusti}, T., {de Bruijne}, J.~H.~J., {et~al.} 2016,
  \aap, 595, A1

\bibitem[{{Ginsburg} {et~al.}(2019){Ginsburg}, {Sip{\H o}cz}, {Brasseur},
  {Cowperthwaite}, {Craig}, {Deil}, {Guillochon}, {Guzman}, {Liedtke}, {Lian
  Lim}, {Lockhart}, {Mommert}, {Morris}, {Norman}, {Parikh}, {Persson},
  {Robitaille}, {Segovia}, {Singer}, {Tollerud}, {de Val-Borro}, {Valtchanov},
  {Woillez}, {The Astroquery collaboration}, \& {a subset of the astropy
  collaboration}}]{2019AJ....157...98G}
{Ginsburg}, A., {Sip{\H o}cz}, B.~M., {Brasseur}, C.~E., {et~al.} 2019, \aj,
  157, 98

\bibitem[{Ginsburg {et~al.}(2024)Ginsburg, Sipőcz, Brasseur, Parikh,
  jcsegovia, Groener, Norman, derdon, Kelley, Robitaille, Lim, Vaher, Deil,
  Mommert, Medina, Tollerud, Nilsson, Baumann, Craig, de~Val-Borro, Weaver,
  jespinosaar, Davies, Adeleke, Cowboy, Persson, Dempsey, syed gilani, Mesh, \&
  Mirocha}]{astroquery_10799414}
Ginsburg, A., Sipőcz, B., Brasseur, C.~E., {et~al.} 2024, astropy/astroquery:
  v0.4.7

\bibitem[{Gommers {et~al.}(2024)Gommers, Virtanen, Haberland, Burovski,
  Weckesser, Reddy, Oliphant, Cournapeau, Nelson, alexbrc, Roy, Peterson,
  Polat, Wilson, endolith, Mayorov, van~der Walt, Brett, Laxalde, Larson,
  Sakai, Millman, Lars, peterbell10, Carey, van Mulbregt, eric jones, McKibben,
  Kai, \& Kern}]{scipy_10909890}
Gommers, R., Virtanen, P., Haberland, M., {et~al.} 2024, scipy/scipy: SciPy
  1.13.0

\bibitem[{{G{\'o}rski} {et~al.}(2005){G{\'o}rski}, {Hivon}, {Banday},
  {Wandelt}, {Hansen}, {Reinecke}, \& {Bartelmann}}]{2005ApJ...622..759G}
{G{\'o}rski}, K.~M., {Hivon}, E., {Banday}, A.~J., {et~al.} 2005, \apj, 622,
  759

\bibitem[{{Green} {et~al.}(2019){Green}, {Schlafly}, {Zucker}, {Speagle}, \&
  {Finkbeiner}}]{2019ApJ...887...93G}
{Green}, G.~M., {Schlafly}, E., {Zucker}, C., {Speagle}, J.~S., \&
  {Finkbeiner}, D. 2019, \apj, 887, 93

\bibitem[{Grisel {et~al.}(2024)Grisel, Mueller, Lars, Gramfort, Louppe, Fan,
  Prettenhofer, Blondel, Niculae, Nothman, Lemaitre, Joly, Estève,
  du~Boisberranger, Vanderplas, manoj kumar, Qin, Hug, Varoquaux, Layton,
  Jalali, (Venkat)~Raghav, Schönberger, Jerphanion, Yurchak, Liu, Lorentzen,
  la~Tour, Li, \& Marmo}]{scikit-learn_10951361}
Grisel, O., Mueller, A., Lars, {et~al.} 2024, scikit-learn/scikit-learn:
  Scikit-learn 1.4.2

\bibitem[{Harris {et~al.}(2020)Harris, Millman, van~der Walt, Gommers,
  Virtanen, Cournapeau, Wieser, Taylor, Berg, Smith, Kern, Picus, Hoyer, van
  Kerkwijk, Brett, Haldane, del R{\'{i}}o, Wiebe, Peterson,
  G{\'{e}}rard-Marchant, Sheppard, Reddy, Weckesser, Abbasi, Gohlke, \&
  Oliphant}]{numpy}
Harris, C.~R., Millman, K.~J., van~der Walt, S.~J., {et~al.} 2020, Nature, 585,
  357

\bibitem[{{He}(2023)}]{He2023warp}
{He}, Z. 2023, \apjl, 954, L9

\bibitem[{{Herschel}(1786)}]{1786RSPT...76..457H}
{Herschel}, W. 1786, Philosophical Transactions of the Royal Society of London
  Series I, 76, 457

\bibitem[{{Herschel}(1789)}]{1789RSPT...79..212H}
{Herschel}, W. 1789, Philosophical Transactions of the Royal Society of London
  Series I, 79, 212

\bibitem[{{Hobbs} \& {H{\o}g}(2018)}]{GaiaNIR2018}
{Hobbs}, D. \& {H{\o}g}, E. 2018, in Astrometry and Astrophysics in the Gaia
  Sky, ed. A.~{Recio-Blanco}, P.~{de Laverny}, A.~G.~A. {Brown}, \&
  T.~{Prusti}, Vol. 330, 67--70

\bibitem[{{Hunt} \& {Reffert}(2021)}]{2021A&A...646A.104H}
{Hunt}, E.~L. \& {Reffert}, S. 2021, \aap, 646, A104

\bibitem[{{Hunt} \& {Reffert}(2023)}]{2023A&A...673A.114H}
{Hunt}, E.~L. \& {Reffert}, S. 2023, \aap, 673, A114

\bibitem[{{Hunt} \& {Reffert}(2024)}]{2024A&A...686A..42H}
{Hunt}, E.~L. \& {Reffert}, S. 2024, \aap, 686, A42

\bibitem[{Hunter(2007)}]{Hunter:2007}
Hunter, J.~D. 2007, Computing in Science \& Engineering, 9, 90

\bibitem[{{Ivezi{\'c}} {et~al.}(2019){Ivezi{\'c}}, {Kahn}, {Tyson}, {Abel},
  {Acosta}, {Allsman}, {Alonso}, {AlSayyad}, {Anderson}, {Andrew}, {Angel},
  {Angeli}, {Ansari}, {Antilogus}, {Araujo}, {Armstrong}, {Arndt}, {Astier},
  {Aubourg}, {Auza}, {Axelrod}, {Bard}, {Barr}, {Barrau}, {Bartlett}, {Bauer},
  {Bauman}, {Baumont}, {Bechtol}, {Bechtol}, {Becker}, {Becla}, {Beldica},
  {Bellavia}, {Bianco}, {Biswas}, {Blanc}, {Blazek}, {Blandford}, {Bloom},
  {Bogart}, {Bond}, {Booth}, {Borgland}, {Borne}, {Bosch}, {Boutigny},
  {Brackett}, {Bradshaw}, {Brandt}, {Brown}, {Bullock}, {Burchat}, {Burke},
  {Cagnoli}, {Calabrese}, {Callahan}, {Callen}, {Carlin}, {Carlson},
  {Chandrasekharan}, {Charles-Emerson}, {Chesley}, {Cheu}, {Chiang}, {Chiang},
  {Chirino}, {Chow}, {Ciardi}, {Claver}, {Cohen-Tanugi}, {Cockrum}, {Coles},
  {Connolly}, {Cook}, {Cooray}, {Covey}, {Cribbs}, {Cui}, {Cutri}, {Daly},
  {Daniel}, {Daruich}, {Daubard}, {Daues}, {Dawson}, {Delgado}, {Dellapenna},
  {de Peyster}, {de Val-Borro}, {Digel}, {Doherty}, {Dubois},
  {Dubois-Felsmann}, {Durech}, {Economou}, {Eifler}, {Eracleous}, {Emmons},
  {Fausti Neto}, {Ferguson}, {Figueroa}, {Fisher-Levine}, {Focke}, {Foss},
  {Frank}, {Freemon}, {Gangler}, {Gawiser}, {Geary}, {Gee}, {Geha}, {Gessner},
  {Gibson}, {Gilmore}, {Glanzman}, {Glick}, {Goldina}, {Goldstein}, {Goodenow},
  {Graham}, {Gressler}, {Gris}, {Guy}, {Guyonnet}, {Haller}, {Harris},
  {Hascall}, {Haupt}, {Hernandez}, {Herrmann}, {Hileman}, {Hoblitt}, {Hodgson},
  {Hogan}, {Howard}, {Huang}, {Huffer}, {Ingraham}, {Innes}, {Jacoby}, {Jain},
  {Jammes}, {Jee}, {Jenness}, {Jernigan}, {Jevremovi{\'c}}, {Johns}, {Johnson},
  {Johnson}, {Jones}, {Juramy-Gilles}, {Juri{\'c}}, {Kalirai}, {Kallivayalil},
  {Kalmbach}, {Kantor}, {Karst}, {Kasliwal}, {Kelly}, {Kessler}, {Kinnison},
  {Kirkby}, {Knox}, {Kotov}, {Krabbendam}, {Krughoff}, {Kub{\'a}nek},
  {Kuczewski}, {Kulkarni}, {Ku}, {Kurita}, {Lage}, {Lambert}, {Lange},
  {Langton}, {Le Guillou}, {Levine}, {Liang}, {Lim}, {Lintott}, {Long},
  {Lopez}, {Lotz}, {Lupton}, {Lust}, {MacArthur}, {Mahabal}, {Mandelbaum},
  {Markiewicz}, {Marsh}, {Marshall}, {Marshall}, {May}, {McKercher}, {McQueen},
  {Meyers}, {Migliore}, {Miller}, {Mills}, {Miraval}, {Moeyens}, {Moolekamp},
  {Monet}, {Moniez}, {Monkewitz}, {Montgomery}, {Morrison}, {Mueller},
  {Muller}, {Mu{\~n}oz Arancibia}, {Neill}, {Newbry}, {Nief}, {Nomerotski},
  {Nordby}, {O'Connor}, {Oliver}, {Olivier}, {Olsen}, {O'Mullane}, {Ortiz},
  {Osier}, {Owen}, {Pain}, {Palecek}, {Parejko}, {Parsons}, {Pease},
  {Peterson}, {Peterson}, {Petravick}, {Libby Petrick}, {Petry},
  {Pierfederici}, {Pietrowicz}, {Pike}, {Pinto}, {Plante}, {Plate}, {Plutchak},
  {Price}, {Prouza}, {Radeka}, {Rajagopal}, {Rasmussen}, {Regnault}, {Reil},
  {Reiss}, {Reuter}, {Ridgway}, {Riot}, {Ritz}, {Robinson}, {Roby}, {Roodman},
  {Rosing}, {Roucelle}, {Rumore}, {Russo}, {Saha}, {Sassolas}, {Schalk},
  {Schellart}, {Schindler}, {Schmidt}, {Schneider}, {Schneider}, {Schoening},
  {Schumacher}, {Schwamb}, {Sebag}, {Selvy}, {Sembroski}, {Seppala}, {Serio},
  {Serrano}, {Shaw}, {Shipsey}, {Sick}, {Silvestri}, {Slater}, {Smith},
  {Smith}, {Sobhani}, {Soldahl}, {Storrie-Lombardi}, {Stover}, {Strauss},
  {Street}, {Stubbs}, {Sullivan}, {Sweeney}, {Swinbank}, {Szalay}, {Takacs},
  {Tether}, {Thaler}, {Thayer}, {Thomas}, {Thornton}, {Thukral}, {Tice},
  {Trilling}, {Turri}, {Van Berg}, {Vanden Berk}, {Vetter}, {Virieux},
  {Vucina}, {Wahl}, {Walkowicz}, {Walsh}, {Walter}, {Wang}, {Wang}, {Warner},
  {Wiecha}, {Willman}, {Winters}, {Wittman}, {Wolff}, {Wood-Vasey}, {Wu},
  {Xin}, {Yoachim}, \& {Zhan}}]{2019ApJ...873..111I}
{Ivezi{\'c}}, {\v{Z}}., {Kahn}, S.~M., {Tyson}, J.~A., {et~al.} 2019, \apj,
  873, 111

\bibitem[{{Janes} \& {Adler}(1982)}]{1982ApJS...49..425J}
{Janes}, K. \& {Adler}, D. 1982, \apjs, 49, 425

\bibitem[{{Kalberla} {et~al.}(2007){Kalberla}, {Dedes}, {Kerp}, \&
  {Haud}}]{2007A&A...469..511K}
{Kalberla}, P.~M.~W., {Dedes}, L., {Kerp}, J., \& {Haud}, U. 2007, \aap, 469,
  511

\bibitem[{{Kharchenko} {et~al.}(2005){Kharchenko}, {Piskunov}, {R{\"o}ser},
  {Schilbach}, \& {Scholz}}]{2005A&A...440..403K}
{Kharchenko}, N.~V., {Piskunov}, A.~E., {R{\"o}ser}, S., {Schilbach}, E., \&
  {Scholz}, R.~D. 2005, \aap, 440, 403

\bibitem[{{Kharchenko} {et~al.}(2013){Kharchenko}, {Piskunov}, {Schilbach},
  {R{\"o}ser}, \& {Scholz}}]{2013A&A...558A..53K}
{Kharchenko}, N.~V., {Piskunov}, A.~E., {Schilbach}, E., {R{\"o}ser}, S., \&
  {Scholz}, R.~D. 2013, \aap, 558, A53

\bibitem[{{King}(1962)}]{1962AJ.....67..471K}
{King}, I. 1962, \aj, 67, 471

\bibitem[{Kluyver {et~al.}(2016)Kluyver, Ragan-Kelley, P{\'e}rez, Granger,
  Bussonnier, Frederic, Kelley, Hamrick, Grout, Corlay,
  {et~al.}}]{kluyver2016jupyter}
Kluyver, T., Ragan-Kelley, B., P{\'e}rez, F., {et~al.} 2016, in ELPUB, 87--90

\bibitem[{{Koposov} {et~al.}(2008){Koposov}, {Belokurov}, {Evans}, {Hewett},
  {Irwin}, {Gilmore}, {Zucker}, {Rix}, {Fellhauer}, {Bell}, \&
  {Glushkova}}]{2008ApJ...686..279K}
{Koposov}, S., {Belokurov}, V., {Evans}, N.~W., {et~al.} 2008, \apj, 686, 279

\bibitem[{{Kroupa}(2001)}]{Kroupa2001_IMF}
{Kroupa}, P. 2001, \mnras, 322, 231

\bibitem[{{Krumholz} {et~al.}(2019){Krumholz}, {McKee}, \&
  {Bland-Hawthorn}}]{KrumholzMcKee_2019}
{Krumholz}, M.~R., {McKee}, C.~F., \& {Bland-Hawthorn}, J. 2019, Annual Review
  of Astronomy and Astrophysics, 57, 227

\bibitem[{Lada \& Lada(2003)}]{lada_embedded_2003}
Lada, C.~J. \& Lada, E.~A. 2003, Annual Review of Astronomy and Astrophysics,
  41, 57

\bibitem[{{Lam} {et~al.}(2015){Lam}, {Pitrou}, \& {Seibert}}]{numba:2015}
{Lam}, S.~K., {Pitrou}, A., \& {Seibert}, S. 2015, in Proc. Second Workshop on
  the LLVM Compiler Infrastructure in HPC, 1--6

\bibitem[{Lam {et~al.}(2024)Lam, stuartarchibald, Pitrou, Florisson, Markall,
  Seibert, Self-Construct, Anderson, Leobas, rjenc29, Collison, luk-f a,
  Bourque, Kaustubh, Meurer, Oliphant, Riasanovsky, Wang, densmirn,
  KrisMinchev, Masella, Pronovost, njwhite, Wieser, Totoni, Seefeld, Grecco,
  Peterson, Virshup, \& G}]{Numba_10839385}
Lam, S.~K., stuartarchibald, Pitrou, A., {et~al.} 2024, numba/numba: Numba
  0.59.1

\bibitem[{{Lamers} {et~al.}(2005){Lamers}, {Gieles}, {Bastian}, {Baumgardt},
  {Kharchenko}, \& {Portegies Zwart}}]{Lamers2005A&A...441..117L}
{Lamers}, H.~J.~G.~L.~M., {Gieles}, M., {Bastian}, N., {et~al.} 2005, \aap,
  441, 117

\bibitem[{{Lemasle} {et~al.}(2022){Lemasle}, {Lala}, {Kovtyukh}, {Hanke},
  {Prudil}, {Bono}, {Braga}, {da Silva}, {Fabrizio}, {Fiorentino},
  {Fran{\c{c}}ois}, {Grebel}, \& {Kniazev}}]{2022A&A...668A..40L}
{Lemasle}, B., {Lala}, H.~N., {Kovtyukh}, V., {et~al.} 2022, \aap, 668, A40

\bibitem[{{Levine} {et~al.}(2006){Levine}, {Blitz}, \&
  {Heiles}}]{2006ApJ...643..881L}
{Levine}, E.~S., {Blitz}, L., \& {Heiles}, C. 2006, \apj, 643, 881

\bibitem[{{Lindegren} \& {Dravins}(2021)}]{Lindegren2021}
{Lindegren}, L. \& {Dravins}, D. 2021, \aap, 652, A45

\bibitem[{{Lindegren} {et~al.}(2021){Lindegren}, {Klioner}, {Hern{\'a}ndez},
  {Bombrun}, {Ramos-Lerate}, {Steidelm{\"u}ller}, {Bastian}, {Biermann}, {de
  Torres}, {Gerlach}, {Geyer}, {Hilger}, {Hobbs}, {Lammers}, {McMillan},
  {Stephenson}, {Casta{\~n}eda}, {Davidson}, {Fabricius}, {Gracia-Abril},
  {Portell}, {Rowell}, {Teyssier}, {Torra}, {Bartolom{\'e}}, {Clotet},
  {Garralda}, {Gonz{\'a}lez-Vidal}, {Torra}, {Abbas}, {Altmann}, {Anglada
  Varela}, {Balaguer-N{\'u}{\~n}ez}, {Balog}, {Barache}, {Becciani}, {Bernet},
  {Bertone}, {Bianchi}, {Bouquillon}, {Brown}, {Bucciarelli}, {Busonero},
  {Butkevich}, {Buzzi}, {Cancelliere}, {Carlucci}, {Charlot}, {Cioni},
  {Crosta}, {Crowley}, {del Peloso}, {del Pozo}, {Drimmel}, {Esquej}, {Fienga},
  {Fraile}, {Gai}, {Garcia-Reinaldos}, {Guerra}, {Hambly}, {Hauser},
  {Jan{\ss}en}, {Jordan}, {Kostrzewa-Rutkowska}, {Lattanzi}, {Liao}, {Licata},
  {Lister}, {L{\"o}ffler}, {Marchant}, {Masip}, {Mignard}, {Mints}, {Molina},
  {Mora}, {Morbidelli}, {Murphy}, {Pagani}, {Panuzzo}, {Pe{\~n}alosa Esteller},
  {Poggio}, {Re Fiorentin}, {Riva}, {Sagrist{\`a} Sell{\'e}s}, {Sanchez
  Gimenez}, {Sarasso}, {Sciacca}, {Siddiqui}, {Smart}, {Souami}, {Spagna},
  {Steele}, {Taris}, {Utrilla}, {van Reeven}, \&
  {Vecchiato}}]{2021A&A...649A...2L}
{Lindegren}, L., {Klioner}, S.~A., {Hern{\'a}ndez}, J., {et~al.} 2021, \aap,
  649, A2

\bibitem[{Lundberg \& Lee(2017)}]{shapley_neurips_2017}
Lundberg, S.~M. \& Lee, S.-I. 2017, in Advances in Neural Information
  Processing Systems 30, ed. I.~Guyon, U.~V. Luxburg, S.~Bengio, H.~Wallach,
  R.~Fergus, S.~Vishwanathan, \& R.~Garnett (Curran Associates, Inc.),
  4765--4774

\bibitem[{{Magrini} {et~al.}(2023){Magrini}, {Viscasillas V{\'a}zquez},
  {Spina}, {Randich}, {Romano}, {Franciosini}, {Recio-Blanco}, {Nordlander},
  {D'Orazi}, {Baratella}, {Smiljanic}, {Dantas}, {Pasquini}, {Spitoni},
  {Casali}, {Van der Swaelmen}, {Bensby}, {Stonkute}, {Feltzing}, {Sacco},
  {Bragaglia}, {Pancino}, {Heiter}, {Biazzo}, {Gilmore}, {Bergemann},
  {Tautvai{\v{s}}ien{\.{e}}}, {Worley}, {Hourihane}, {Gonneau}, \&
  {Morbidelli}}]{Magrini2023}
{Magrini}, L., {Viscasillas V{\'a}zquez}, C., {Spina}, L., {et~al.} 2023, \aap,
  669, A119

\bibitem[{McInnes {et~al.}(2017)McInnes, Healy, \& Astels}]{McInnes2017}
McInnes, L., Healy, J., \& Astels, S. 2017, The Journal of Open Source
  Software, 2

\bibitem[{{Messier}(1781)}]{1781cote.rept..227M}
{Messier}, C. 1781, {Catalogue des N{\'e}buleuses et des Amas d'{\'E}toiles
  (Catalog of Nebulae and Star Clusters)}, Connoissance des Temps ou des
  Mouvements C{\'e}lestes, for 1784, p. 227-267

\bibitem[{{Moe} \& {Di Stefano}(2017)}]{2017ApJS..230...15M}
{Moe}, M. \& {Di Stefano}, R. 2017, \apjs, 230, 15

\bibitem[{{Moitinho} {et~al.}(2006){Moitinho}, {V{\'a}zquez}, {Carraro},
  {Baume}, {Giorgi}, \& {Lyra}}]{MoitinhoVazquez_2006}
{Moitinho}, A., {V{\'a}zquez}, R.~A., {Carraro}, G., {et~al.} 2006, Monthly
  Notices of the Royal Astronomical Society, 368, L77

\bibitem[{{Momany} {et~al.}(2006){Momany}, {Zaggia}, {Gilmore}, {Piotto},
  {Carraro}, {Bedin}, \& {de Angeli}}]{MomanyZaggia_2006}
{Momany}, Y., {Zaggia}, S., {Gilmore}, G., {et~al.} 2006, Astronomy and
  Astrophysics, 451, 515

\bibitem[{{Moreira} {et~al.}(2025){Moreira}, {Moitinho}, {Silva}, \&
  {Almeida}}]{Moreira2025A&A...694A..70M}
{Moreira}, S., {Moitinho}, A., {Silva}, A., \& {Almeida}, D. 2025, \aap, 694,
  A70

\bibitem[{{Myers} {et~al.}(2022){Myers}, {Donor}, {Spoo}, {Frinchaboy},
  {Cunha}, {Price-Whelan}, {Majewski}, {Beaton}, {Zasowski}, {O'Connell},
  {Ray}, {Bizyaev}, {Chiappini}, {Garc{\'\i}a-Hern{\'a}ndez}, {Geisler},
  {J{\"o}nsson}, {Lane}, {Longa-Pe{\~n}a}, {Minchev}, {Minniti}, {Nitschelm},
  \& {Roman-Lopes}}]{Myers2022AJ....164...85M}
{Myers}, N., {Donor}, J., {Spoo}, T., {et~al.} 2022, \aj, 164, 85

\bibitem[{{Netopil} {et~al.}(2022){Netopil}, {Oralhan}, {{\c{C}}akmak},
  {Michel}, \& {Karata{\c{s}}}}]{Netopil2022MNRAS.509..421N}
{Netopil}, M., {Oralhan}, {\.I}.~A., {{\c{C}}akmak}, H., {Michel}, R., \&
  {Karata{\c{s}}}, Y. 2022, \mnras, 509, 421

\bibitem[{{Netopil} {et~al.}(2012){Netopil}, {Paunzen}, \&
  {St{\"u}tz}}]{2012ASSP...29...53N}
{Netopil}, M., {Paunzen}, E., \& {St{\"u}tz}, C. 2012, in Astrophysics and
  Space Science Proceedings, Vol.~29, Star Clusters in the Era of Large
  Surveys, 53

\bibitem[{pandas~development team(2024)}]{pandas_10957263}
pandas~development team, T. 2024, pandas-dev/pandas: Pandas

\bibitem[{Pedregosa {et~al.}(2011)Pedregosa, Varoquaux, Gramfort, Michel,
  Thirion, Grisel, Blondel, Prettenhofer, Weiss, Dubourg, Vanderplas, Passos,
  Cournapeau, Brucher, Perrot, \& Duchesnay}]{scikit-learn}
Pedregosa, F., Varoquaux, G., Gramfort, A., {et~al.} 2011, Journal of Machine
  Learning Research, 12, 2825

\bibitem[{{Perez} \& {Granger}(2007)}]{2007CSE.....9c..21P}
{Perez}, F. \& {Granger}, B.~E. 2007, Computing in Science and Engineering, 9,
  21

\bibitem[{{Perren} {et~al.}(2022){Perren}, {Pera}, {Navone}, \&
  {V{\'a}zquez}}]{PerrenDistantOCs2022}
{Perren}, G.~I., {Pera}, M.~S., {Navone}, H.~D., \& {V{\'a}zquez}, R.~A. 2022,
  \aap, 663, A131

\bibitem[{{Perren} {et~al.}(2023){Perren}, {Pera}, {Navone}, \&
  {V{\'a}zquez}}]{2023MNRAS.526.4107P}
{Perren}, G.~I., {Pera}, M.~S., {Navone}, H.~D., \& {V{\'a}zquez}, R.~A. 2023,
  \mnras, 526, 4107

\bibitem[{Petroff(2021)}]{petroff_accessible_color_2021}
Petroff, M.~A. 2021, arXiv e-prints, 2107.02270

\bibitem[{{Pflamm-Altenburg} {et~al.}(2013){Pflamm-Altenburg},
  {Gonz{\'a}lez-L{\'o}pezlira}, \& {Kroupa}}]{2013MNRAS.435.2604P}
{Pflamm-Altenburg}, J., {Gonz{\'a}lez-L{\'o}pezlira}, R.~A., \& {Kroupa}, P.
  2013, \mnras, 435, 2604

\bibitem[{{Pflamm-Altenburg} \& {Kroupa}(2008)}]{2008Natur.455..641P}
{Pflamm-Altenburg}, J. \& {Kroupa}, P. 2008, \nat, 455, 641

\bibitem[{{Portegies Zwart} {et~al.}(2010){Portegies Zwart}, {McMillan}, \&
  {Gieles}}]{2010ARAA..48..431P}
{Portegies Zwart}, S.~F., {McMillan}, S. L.~W., \& {Gieles}, M. 2010, \araa,
  48, 431

\bibitem[{{Riello} {et~al.}(2021){Riello}, {De Angeli}, {Evans}, {Montegriffo},
  {Carrasco}, {Busso}, {Palaversa}, {Burgess}, {Diener}, {Davidson}, {Rowell},
  {Fabricius}, {Jordi}, {Bellazzini}, {Pancino}, {Harrison}, {Cacciari}, {van
  Leeuwen}, {Hambly}, {Hodgkin}, {Osborne}, {Altavilla}, {Barstow}, {Brown},
  {Castellani}, {Cowell}, {De Luise}, {Gilmore}, {Giuffrida}, {Hidalgo},
  {Holland}, {Marinoni}, {Pagani}, {Piersimoni}, {Pulone}, {Ragaini}, {Rainer},
  {Richards}, {Sanna}, {Walton}, {Weiler}, \& {Yoldas}}]{2021A&A...649A...3R}
{Riello}, M., {De Angeli}, F., {Evans}, D.~W., {et~al.} 2021, \aap, 649, A3

\bibitem[{{Rix} {et~al.}(2021){Rix}, {Hogg}, {Boubert}, {Brown}, {Casey},
  {Drimmel}, {Everall}, {Fouesneau}, \& {Price-Whelan}}]{2021AJ....162..142R}
{Rix}, H.-W., {Hogg}, D.~W., {Boubert}, D., {et~al.} 2021, \aj, 162, 142

\bibitem[{{Rybizki} {et~al.}(2022){Rybizki}, {Green}, {Rix}, {El-Badry},
  {Demleitner}, {Zari}, {Udalski}, {Smart}, \& {Gould}}]{2022MNRAS.510.2597R}
{Rybizki}, J., {Green}, G.~M., {Rix}, H.-W., {et~al.} 2022, \mnras, 510, 2597

\bibitem[{{Skowron} {et~al.}(2019){Skowron}, {Skowron}, {Mr{\'o}z}, {Udalski},
  {Pietrukowicz}, {Soszy{\'n}ski}, {Szyma{\'n}ski}, {Poleski}, {Koz{\l}owski},
  {Ulaczyk}, {Rybicki}, \& {Iwanek}}]{Skowron2019Sci...365..478S}
{Skowron}, D.~M., {Skowron}, J., {Mr{\'o}z}, P., {et~al.} 2019, Science, 365,
  478

\bibitem[{{Spina} {et~al.}(2022){Spina}, {Magrini}, \& {Cunha}}]{Spina2022}
{Spina}, L., {Magrini}, L., \& {Cunha}, K. 2022, Universe, 8, 87

\bibitem[{{Sun} {et~al.}(2016){Sun}, {de Grijs}, {Fan}, \&
  {Cameron}}]{2016ApJ...816....9S}
{Sun}, W., {de Grijs}, R., {Fan}, Z., \& {Cameron}, E. 2016, \apj, 816, 9

\bibitem[{{Tang} {et~al.}(2014){Tang}, {Bressan}, {Rosenfield}, {Slemer},
  {Marigo}, {Girardi}, \& {Bianchi}}]{Tang2014MNRAS.445.4287T}
{Tang}, J., {Bressan}, A., {Rosenfield}, P., {et~al.} 2014, \mnras, 445, 4287

\bibitem[{{Tarricq} {et~al.}(2021){Tarricq}, {Soubiran}, {Casamiquela},
  {Cantat-Gaudin}, {Chemin}, {Anders}, {Antoja}, {Romero-G{\'o}mez},
  {Figueras}, {Jordi}, {Bragaglia}, {Balaguer-N{\'u}{\~n}ez}, {Carrera},
  {Castro-Ginard}, {Moitinho}, {Ramos}, \& {Bossini}}]{2021A&A...647A..19T}
{Tarricq}, Y., {Soubiran}, C., {Casamiquela}, L., {et~al.} 2021, \aap, 647, A19

\bibitem[{{Trumpler}(1930)}]{1930LicOB..14..154T}
{Trumpler}, R.~J. 1930, Lick Observatory Bulletin, 420, 154

\bibitem[{{Usher} {et~al.}(2023){Usher}, {Dage}, {Girardi}, {Barmby},
  {Bonatto}, {Chies-Santos}, {Clarkson}, {G{\'o}mez Camus}, {Hartmann},
  {Ferguson}, {Pieres}, {Prisinzano}, {Rhode}, {Rich}, {Ripepi}, {Santiago},
  {Stassun}, {Street}, {Szab{\'o}}, {Venuti}, {Zaggia}, {Canossa}, {Floriano},
  {Lopes}, {Miranda}, {Oliveira}, {Reina-Campos}, {Roman-Lopes}, \&
  {Sobeck}}]{2023PASP..135g4201U}
{Usher}, C., {Dage}, K.~C., {Girardi}, L., {et~al.} 2023, \pasp, 135, 074201

\bibitem[{Van~Rossum \& Drake(2009)}]{python}
Van~Rossum, G. \& Drake, F.~L. 2009, Python 3 Reference Manual (Scotts Valley,
  CA: CreateSpace)

\bibitem[{{V{\'a}zquez} {et~al.}(2008){V{\'a}zquez}, {May}, {Carraro},
  {Bronfman}, {Moitinho}, \& {Baume}}]{VazquezMay_2008}
{V{\'a}zquez}, R.~A., {May}, J., {Carraro}, G., {et~al.} 2008, The
  Astrophysical Journal, 672, 930

\bibitem[{Virtanen {et~al.}(2020)Virtanen, Gommers, Oliphant, Haberland, Reddy,
  Cournapeau, Burovski, Peterson, Weckesser, Bright, {van der Walt}, Brett,
  Wilson, Millman, Mayorov, Nelson, Jones, Kern, Larson, Carey, Polat, Feng,
  Moore, {VanderPlas}, Laxalde, Perktold, Cimrman, Henriksen, Quintero, Harris,
  Archibald, Ribeiro, Pedregosa, {van Mulbregt}, \& {SciPy 1.0
  Contributors}}]{2020SciPy-NMeth}
Virtanen, P., Gommers, R., Oliphant, T.~E., {et~al.} 2020, Nature Methods, 17,
  261

\bibitem[{{Viscasillas V{\'a}zquez} {et~al.}(2023){Viscasillas V{\'a}zquez},
  {Magrini}, {Spina}, {Tautvai{\v{s}}ien{\.{e}}}, {Van der Swaelmen},
  {Randich}, \& {Sacco}}]{Viscasillas2023A&A...679A.122V}
{Viscasillas V{\'a}zquez}, C., {Magrini}, L., {Spina}, L., {et~al.} 2023, \aap,
  679, A122

\bibitem[{Wagg {et~al.}(2024)Wagg, Broekgaarden, \&
  Gültekin}]{software-citation-station-zenodo}
Wagg, T., Broekgaarden, F., \& Gültekin, K. 2024,
  TomWagg/software-citation-station: v1.2

\bibitem[{{Wagg} \& {Broekgaarden}(2024)}]{software-citation-station-paper}
{Wagg}, T. \& {Broekgaarden}, F.~S. 2024, arXiv e-prints, arXiv:2406.04405

\bibitem[{{Wenger, M.} {et~al.}(2000){Wenger, M.}, {Ochsenbein, F.}, {Egret,
  D.}, {Dubois, P.}, {Bonnarel, F.}, {Borde, S.}, {Genova, F.}, {Jasniewicz,
  G.}, {Lalo{\"e}, S.}, {Lesteven, S.}, \& {Monier,
  R.}}]{wengerm_simbad_astronomical_2000}
{Wenger, M.}, {Ochsenbein, F.}, {Egret, D.}, {et~al.} 2000, Astron. Astrophys.
  Suppl. Ser., 143, 9

\bibitem[{{W}es {M}c{K}inney(2010)}]{mckinney-proc-scipy-2010}
{W}es {M}c{K}inney. 2010, in {P}roceedings of the 9th {P}ython in {S}cience
  {C}onference, ed. {S}t\'efan van~der {W}alt \& {J}arrod {M}illman, 56 -- 61

\bibitem[{{Zhang} {et~al.}(2021){Zhang}, {Chen}, \&
  {Zhao}}]{Zhang2021ApJ...919...52Z}
{Zhang}, H., {Chen}, Y., \& {Zhao}, G. 2021, \apj, 919, 52

\end{thebibliography}

\appendix

\onecolumn

\section{Correlations in detectability between cluster parameters}

\begin{figure}[h!]
    \centering    
    \includegraphics[trim=3.5cm 3.5cm 3.cm 5cm, clip,width=\textwidth]{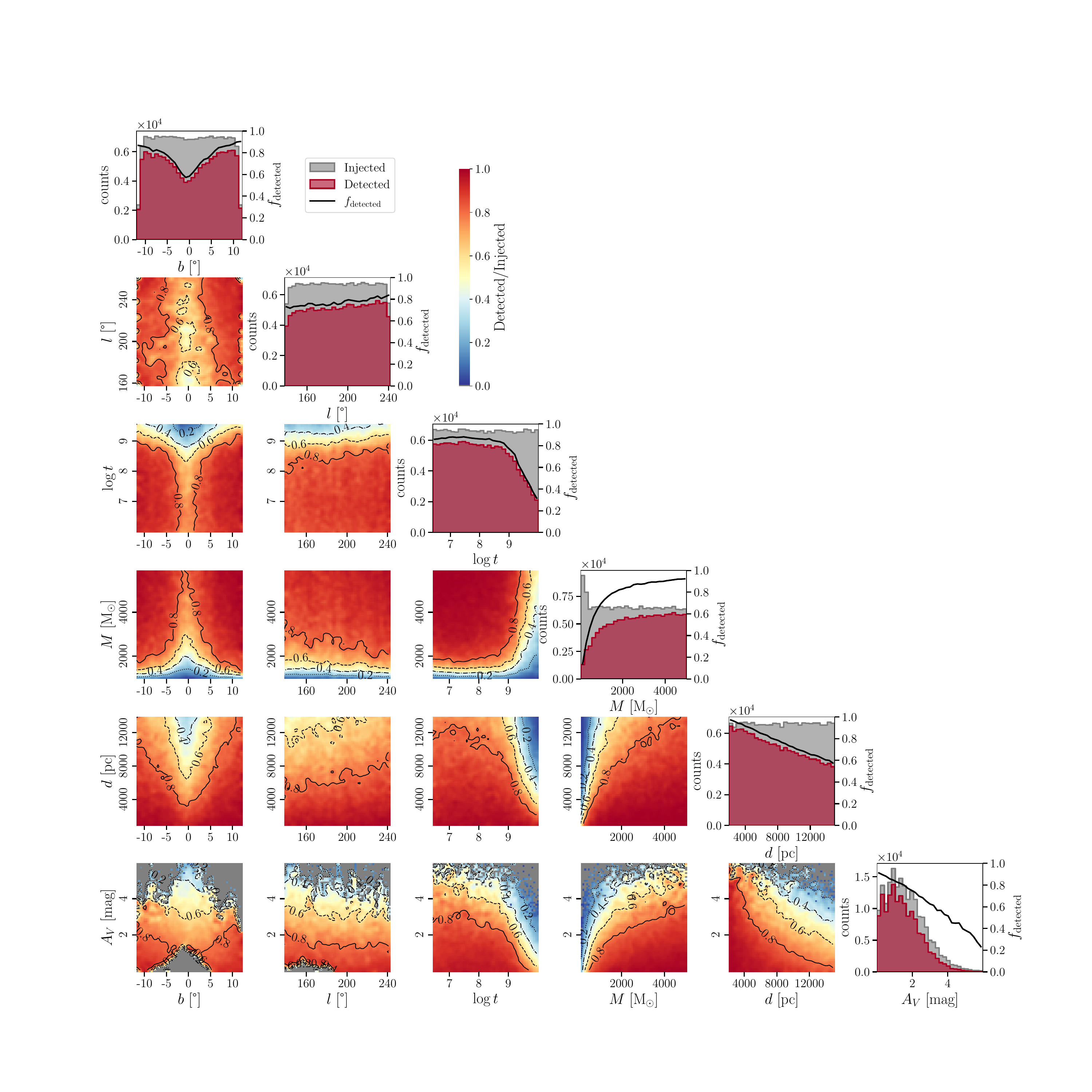} 
    \caption{\label{fig:detection_trends_cornerplot}Cornerplot summarising the covariances between parameters further to those discussed in Sect.~\ref{sec:results_trends} and shown in Fig.~\ref{fig:detection_trends_grid}. In the diagonal panels, we show the 1D parameter distributions of the simulated and recovered clusters in grey and red respectively. The fraction of detected objects are shown as a solid black line. The off-diagonal panels are heat maps showing the fraction of injected clusters that were recovered as a function of two parameters. We superimpose contours of iso-detectability with dotted, dash-dotted, dashed, and solid lines at $f_{\rm detected}$ equal to $0.2$, $0.4$, $0.6$, and $0.8$, respectively. Grey areas indicate regions of the parameter space where we do not inject OCs. These individual heatmaps can be used to visually find regions where, for instance, almost all clusters are detectable (solid red regions), where almost no clusters are detectable (solid blue regions), or regions where detectability changes sharply as a function of certain parameters, as well as the trend between them.}
\end{figure}

\newpage
\twocolumn

\section{Injected cluster proper motions}

\begin{figure}[h]
    \includegraphics[width=\columnwidth]{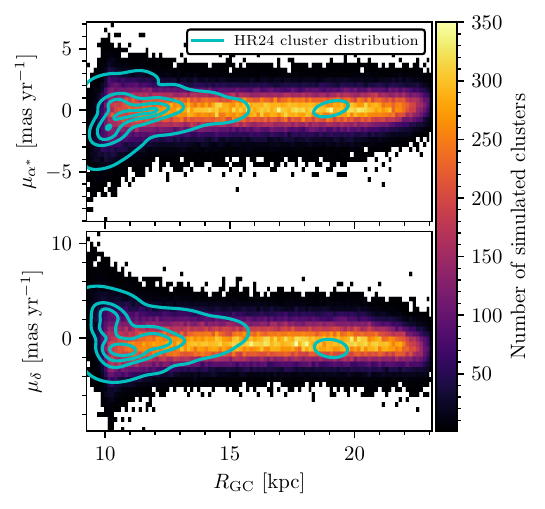} 
    \caption{\label{fig:proper_motions} Comparison between the proper motions of our injected clusters and clusters in HR24 as a function of galactocentric radius. All 192\,318 simulated clusters are shown by the heatmap, with a colour bar on the right. The proper motion distribution of the 818 HR24 clusters within our studied region is shown by the kernel density estimate, with contours at the 0.1\%, 20\%, 50\%, and 80\% density levels.}
\end{figure}

\section{Interpreting the machine learning model}\label{app:interpretation}

\begin{figure}
    \includegraphics[width=\columnwidth]{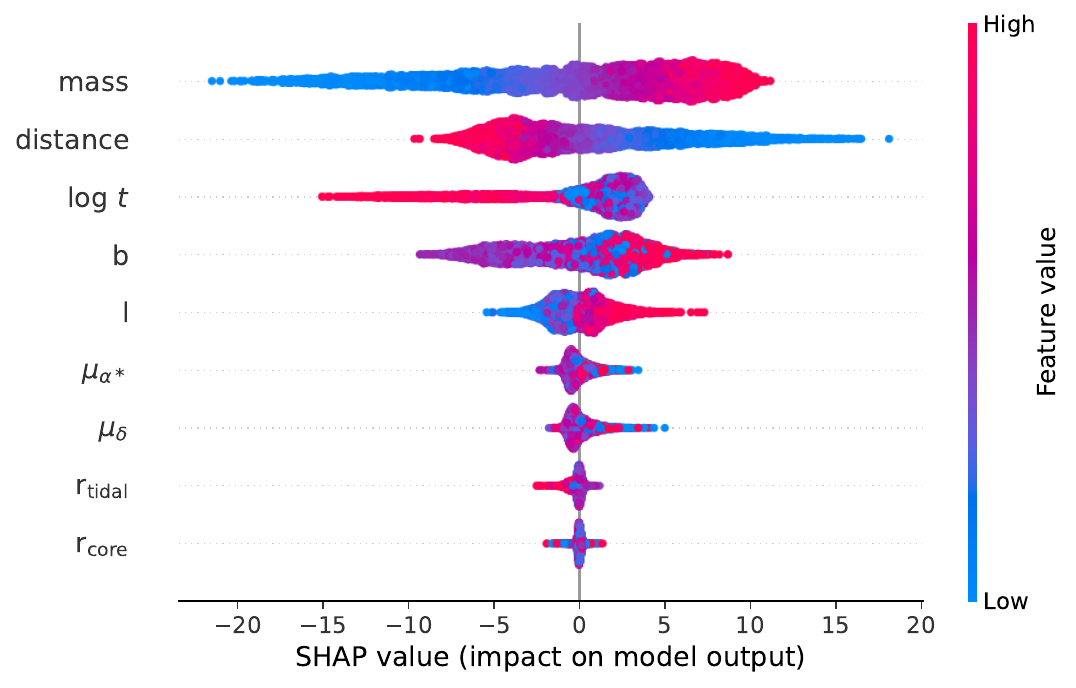} 
    \includegraphics[width=\columnwidth]{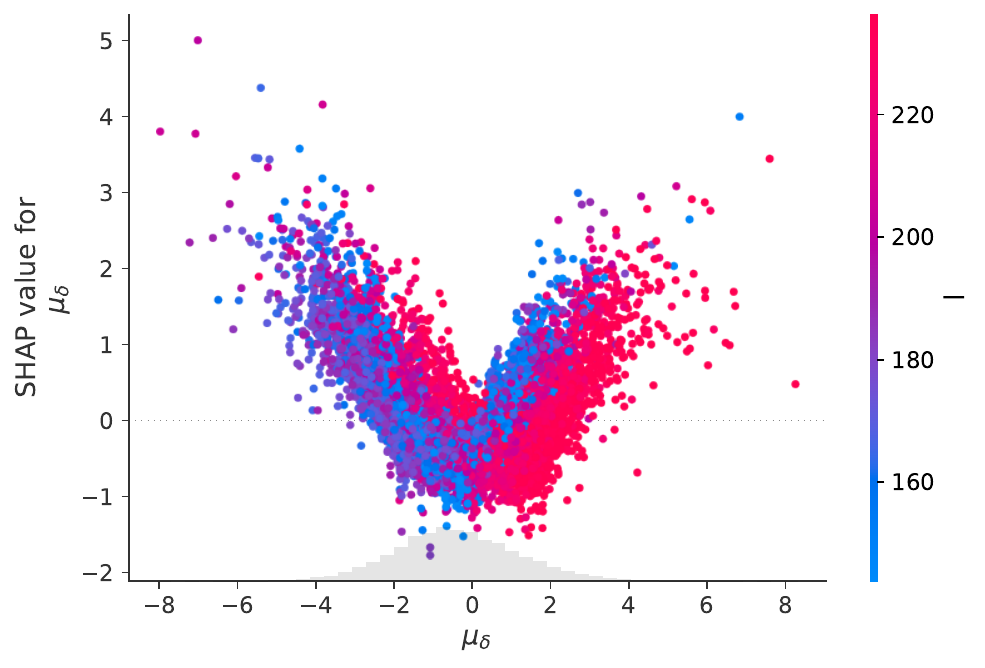} 
    \caption{\label{fig:cst_shap} SHAP feature importance values for the CST predictor. \emph{Top:} Beeswarm plot where each cluster in the validation dataset is shown as a dot. Each row corresponds to the impact of a different input parameter. The colour coding corresponds to whether it was a high or low value of the parameter. The $x$ axis shows the final impact on the output of the model, which is how much the CST is changed for that given cluster and that given parameter value. For example: for cluster mass, low mass values (blue) correspond to a much lower SHAP/CST, whereas high mass values (red) correspond to a much higher SHAP/CST. On the other hand, most age values have minimal impact on CST, although high ages significantly reduce it. \emph{Bottom:} SHAP value at a given \texttt{pmdec} as a function of \texttt{pmdec} and shown for all clusters in the validation dataset. Colour coding shows the Galactic longitude, $l$.}
\end{figure}

To better quantify why certain clusters can or cannot be detected, we applied Shapley Additive Explanations\footnote{\url{https://github.com/shap/shap}} \citep[SHAP][]{shapley_neurips_2017} to explain the importance of features to our CST prediction model from Sect.~\ref{sec:cst_fdetected_predictor}. Figure~\ref{fig:cst_shap} shows the SHAP feature importance of different cluster parameters. For the upper panel in Fig.~\ref{fig:cst_shap}, the width and colour coding of each row of the beeswarm plot shows the impact of different parameters on the CST. Cluster mass, distance, age, and $b$ have the largest impact on cluster significance respectively. The $x$ axis shows the impact of different input features on the CST that the model predicts, and the difference in CST between the upper and lower values in our studied parameter ranges from Table~\ref{tab:parameter_choices} can hence be read directly off of the $x$ axis of the plot. For instance, a cluster with a mass of 50~M\textsubscript{\sun} will have a CST 30$\sigma$ lower than a cluster with a mass of 5000~M\textsubscript{\sun}.

The SHAP interpretation also allows for the study of more subtle parameter effects. Although proper motion has a smaller impact than other parameters on the detectability of a cluster, trends in how it impacts CST are still visible. The bottom panel of Fig.~\ref{fig:cst_shap} shows how the CST of the model changes as a function of $\mu_\delta$ and at given Galactic longitude values. The impact of cluster proper motion on CST is a `V' shape, with values of proper motion distinct from that of the field causing cluster detectability to increase. As $l$ increases, the typical proper motion of the field also increases slightly, causing the `V' shape to traverse to higher $\mu_\delta$ values. In general, proper motion at the distance range studied in this work causes CST to increase by around 3$\sigma$ for every 4~mas\,yr\textsuperscript{-1} that the cluster's $\mu_\delta$ is different from that of the field. A CST change of 3$\sigma$ could still be the difference between a cluster being detected and included in HR23 or not. This also shows that studies of the velocity distribution of OCs based on \emph{Gaia} data will be biased towards including clusters with more extreme orbits unless they also account for selection effects.

\end{document}